\documentclass{aa}

\DeclareMathAlphabet{\altmathcal}{OMS}{cmsy}{m}{n}
\usepackage{mathptmx}

\usepackage[T1]{fontenc}
\usepackage{txfonts}

\usepackage{graphicx}	
\usepackage{amsmath}	
\usepackage{amssymb}	
\usepackage{gensymb}
\usepackage{xcolor}     
\usepackage{cancel}
\usepackage{multirow}
\usepackage{mathrsfs}
\usepackage{ulem}
\usepackage[hidelinks,colorlinks=true,linkcolor=blue,citecolor=blue]{hyperref}

\renewcommand{\vec}[1]{\mathbf{#1}}
\renewcommand{\tens}[1]{\mathsf{#1}}

\newcommand{\pd}[2]{\frac{\partial #1}{\partial #2} }

\newcommand{\DS}{\displaystyle}

\newcommand{\cR}{\altmathcal{R}}
\newcommand{\cS}{\altmathcal{S}}

\newcommand{\cU}{\altmathcal{U}}

\newcommand{\RED}{\color{red}}

\DeclareMathOperator{\sign}{sign}

\begin{document}

\title{Resistive relativistic MHD simulations of astrophysical jets}

\author{G. Mattia      \inst{1}
\and    L. Del Zanna   \inst{2,1,3}
\and    M. Bugli       \inst{4,5,6}
\and    A. Pavan       \inst{7,8,9}
\and    R. Ciolfi      \inst{8,9}
\and    G. Bodo        \inst{10}
\and    A. Mignone     \inst{4,6}
}

\institute{INFN , Sezione di Firenze, Via G. Sansone 1, I-50019 Sesto Fiorentino (FI), Italy \\ \email{mattia@fi.infn.it}
\and        Dipartimento di Fisica e Astronomia, Universit\`a di Firenze, Via G. Sansone 1, I-50019 Sesto Fiorentino (FI), Italy
\and        INAF, Osservatorio Astrofisico di Arcetri, Largo E. Fermi 5, I-50125 Firenze, Italy
\and        Dipartimento di Fisica, Universit\`a di Torino, Via P. Giuria 1, I-10125 Torino, Italy
\and        Universit\'e Paris-Saclay, Universit\'e Paris Cit\'e, CEA, CNRS, AIM, F-91191, Gif-sur-Yvette, France
\and        INFN, Sezione di Torino, Via P. Giuria 1, I-10125 Torino, Italy
\and        Dipartimento di Fisica e Astronomia, Universit\`a di Padova, Via F. Marzolo 8, I-35131 Padova, Italy
\and        INAF, Osservatorio Astronomico di Padova, Vicolo dell'Osservatorio 5, I-35122 Padova, Italy
\and        INFN, Sezione di Padova, Via F. Marzolo 8, I-35131 Padova, Italy
\and        INAF, Osservatorio Astrofisico di Torino, Strada Osservatorio 20, I-10025 Pino Torinese (TO), Italy
}

\date{Received XXX; accepted YYY}

\abstract
{} 
{The main goal of the present paper is to provide the first systematic numerical study of the propagation
of astrophysical relativistic jets, in the context of high-resolution shock-capturing resistive relativistic 
magnetohydrodynamics (RRMHD) simulations. 
We aim at investigating different values and models for the plasma resistivity coefficient, and at assessing their impact on the level of turbulence, the formation of current sheets and reconnection plasmoids, the electromagnetic energy content, and the dissipated power.}
{We use the PLUTO code for simulations and we assume an axisymmetric setup for jets, endowed with both poloidal and toroidal magnetic fields, and propagating in a uniform magnetized medium. 
The gas is assumed to be characterized by a realistic Synge-like equation of state (Taub equation), appropriate for such type of astrophysical jets. 
The Taub equation is combined here for the first time with the {\it Implicit-Explicit Runge-Kutta} time-stepping
procedure, as required in RRMHD simulations.}
{
The main result is that turbulence is clearly suppressed for the highest values of resistivity (low Lundquist numbers), current sheets are broader, and plasmoids are barely present, while for low values of resistivity results are very similar to ideal runs, where dissipation is purely numerical.
We find that recipes employing a variable resistivity based on the advection of a jet tracer or on the assumption of a uniform Lundquist number improve on the use of a constant coefficient and are probably more realistic, preserving the development of turbulence and of sharp current sheets, possible sites for the acceleration of the non-thermal particles producing the observed high-energy emission.}
{}

              
\maketitle



\section{Introduction}
%
%
%

Astrophysical jets, i.e. supersonic collimated outflows, represent a ubiquitous phenomenon in the Universe, 
characterizing various classes of celestial sources at very different spatial scales and evolutionary stages.
In particular, the interplay of strong gravity, rotation, and magnetic fields of compact objects may cause
the launching of relativistic jets, even with high Lorentz factors, as spectacularly demonstrated by their
apparent superluminal motion when propagation is close to the line of sight, and more recently
by the imaging of the supermassive black hole as the inner engine of the kiloparsec-scale jet of M87 \citep{EHT2019}.
Other than in Active Galactic Nuclei (AGN), relativistic outflows and jets are also present in several types 
of stellar-sized astrophysical bodies, such as the collapsing sources of long Gamma-Ray Bursts (GRB) 
\citep{Woosley1993}, Pulsar Wind Nebulae (PWN) \citep{Weisskopf2000}, and recently we had the 
first multi-messenger indication that also Binary Neutron Star (BNS) merger events are able 
to drive relativistic outflows in the form of short GRBs \citep{Abbott2017}.
The electromagnetic spectral signature of such objects often features highly non-thermal radiation, mostly 
synchrotron and inverse Compton processes, due to accelerated relativistic particles (electrons).
In the case of AGNs, particles may be accelerated either in the inner engine and advected, or along the jet 
and at the terminal radio lobes (e.g. \citealt{BBR1984,BMR2019}).
More recently, even protostellar jets have shown similar radiation features \citep{Leeetal2018}.

Different processes have been invoked in order to explain such particles acceleration in astrophysical sources.
In particular, relativistic magnetic (fast) reconnection, i.e. the impulsive topological rearrangement of
field lines in a magnetically dominated or very hot plasma  \citep{LyuUdz2003,Lyubarsky2005,KBL2007,DzPLBB2016,RPSK2019} is a very general and promising acceleration process, 
which has been extensively investigated in recent years, especially through particle-in-cell (PIC) simulations \citep{SirSpi2014,Cerutti2014,ComSir2018,ZSG2021}.
The precise mechanism regulating the acceleration of particles during the reconnection process, namely
the formation of X and O points and the subsequent merging of plasmoids \citep[e.g.][]{LouUdz2016},
is still debated, as well as their impact on the large-scale scenario \citep[e.g.][]{Medetal2023}.

Because of the unavoidable numerical resistivity present in every MHD simulation (even within the ideal framework), the disentanglement of the effects due to the truncation error from the physical process is a key point in order to understand the role of magnetic reconnection in astrophysical jets.
The most critical issue of the ideal approximation (which constraints the electric field to the standard $-\vec{v}\times\vec{B}$) is the lack of control on the numerical diffusivity, which becomes strongly susceptible on the grid resolution and numerical algorithms \citep{Nunez-delarosa2016,Felker2018}.

It is therefore important to perform numerical simulations of astrophysical plasmas going beyond the usual assumption of ideal MHD, almost invariably employed especially in the relativistic regime.
The simplest possible framework in which the role of a finite plasma conductivity can be investigated is that of the resistive relativistic magnetohydrodynamics (RRMHD) regime, in which the electric field in the coming frame is not vanishing as in the ideal MHD case but is simply proportional to the local current density through a (scalar for simplicity) resistivity coefficient $\eta$. 
By doing so, the electric field in the laboratory frame is allowed to depart from its ideal configuration (which imposes a vanishing component along the magnetic field lines) especially in current sheets, where reconnection events and particle acceleration are expected to take place.
Obviously any current numerical modeling will never achieve the realistic low values of $\eta$ typical of \textit{collisionless} astrophysical plasmas, though it is well known that small-scale unresolved turbulent fluctuations may act as sources of additional dissipation (see the Conclusions for additional comments on sub-grid modeling).

Here we just want to stress that the RRMHD regime allows to single out the regions where resistive effects may be important, by adopting a physically motivated model. This is instead impossible in numerical ideal MHD, where magnetic dissipation is invariably present at the grid level (due to numerical round-off errors), and any non-ideal aspect cannot be controlled.
Moreover, employing an explicit resistivity allows for more meaningful comparisons between models produced with different codes and numerical algorithms, as the uncertainties related to magnetic numerical dissipation are removed.
Thus, this approach promotes the reproducibility of jet simulations and strengthens the physical interpretation of their results.

Most of the RRMHD numerical investigations of reconnection, either due to the tearing instability
in pre-existing current sheets (e.g. \citealt{DzPLBB2016}), or formed in the process of a turbulent cascade
(e.g. \citealt{CRP2021}), involve localized portions of plasma and periodic
boundary conditions. Other studies assume reconnection induced by current-driven kink instabilities in
(periodic) columns of magnetized plasma \citep{Medetal2021,BTS2021}.
A few investigations are concerned with the accretion disk and launching region at the base of the jet, some of them also addressing the problem of the flaring activity observed in Sgr A$^*$ \citep{QFNB2017,VFQN2019,RPA2020,NMPFR2022,RLCetal2022}, though the last works assume \textit{ideal} conditions.
Other numerical investigations combine RRMHD with mean-field dynamo action in order to explain the growth of the magnetic field in accretion disks up to the required equipartition values \citep{TDzBB2020,TDzBB2021,DzTFBB2022}.

On the other hand, to our knowledge there are no studies of 
{\it propagating} jets using RRMHD simulations with finite plasma conductivity.
The aim of the present work is hence precisely to perform the first systematic numerical investigation 
of propagating magnetized jets via RRMHD simulations. 
The combination of high-Mach number collimated ouflows, strong magnetization, and turbulence development 
is particularly challenging in non-ideal relativistic MHD. Robust schemes and high resolution are both required, 
and this must be the reason why such effort has not been faced yet,
to our knowledge. 

In the present work we perform and discuss high-resolution axisymmetric simulations
(in cylindrical coordinates and flat Minkowskian metric) of jets endowed with both poloidal and toroidal magnetic fields, propagating in a uniform magnetized medium. 
Simulations are obtained with the PLUTO\footnote{http://plutocode.ph.unito.it/} shock-capturing code \citep{PLUTO2007,PLUTO2012}.
The gas is assumed to be characterized by a realistic Synge-like equation of state (Taub equation), 
believed to be appropriate for such type astrophysical jets \citep{MigMck2007}.
The Taub equation of state is combined here for the first time with {\it Implicit-Explicit Runge-Kutta} \citep{ParRus2005,PLRR2009}
routines for time-stepping, which allow a proper treatment of stiff terms in the evolutionary equation
for the electric field in the limit of small resistivity coefficients.

Various levels of magnetization and plasma beta are investigated, and different values and models for the
resistivity coefficient (assumed to be a scalar, but not necessarily a constant) are proposed, taking care that numerical diffusivity remains lower than the physical one.
Results are compared, in terms of morphology and turbulence level in the jet. Regions of high values of the current density and the possible sites for particle acceleration inside current sheets are singled out, zooming promising regions with evolving plasmoids in selected cases.
In order to quantify and characterize the role of resistivity, we then show important quantities like the electromagnetic energy content in the jet, the dissipated Ohmic power, and the non-ideal component of the electric field, which is no longer perpendicular to both the magnetic field and the velocity in the resistive case.
We show that our combination of numerical algorithms is able to ensure robustness for the
variety of the explored parameters, and highly accurate results, in terms of small-scale turbulent
features and reconnection sites. All results are reported in Section 3, whereas Section 2 is devoted 
to the illustration of the system of equations solved and the numerical recipes employed, and
conclusions and future applications will be discussed in Section 4. Numerical benchmarks, resolution tests and comparisons with other numerical schemes and with ideal simulations are reported in the appendices. 
\section{Equations and numerical methods}
\label{sec::Equations}
%
%
%

In this section, we briefly describe the fundamental equations of resistive relativistic MHD \citep{Anile2005,Komissarov2007}, first in covariant form and then specialized to Minkowskian flat space, splitting time and space components.
We then shortly summarize the numerical algorithms employed for this work.

\subsection{The system of Resistive Relativistic MHD (RRMHD)}
\label{sec::RRMHD_Eq}
%

The equations adopted in order to describe resistive relativistic plasmas, consist in the Maxwell equations:
\begin{equation}
\nabla_\mu F^{\mu\nu}  = -J^\nu , \qquad\qquad
\nabla_\mu F^{*\mu\nu} = 0,
\end{equation}
where $F^{\mu\nu}$ and $F^{*\mu\nu}$ are, respectively, the Faraday tensor and its dual, and $J^\mu$ is the four-current density, coupled with a set of conservation laws for, respectively, the mass and (total) energy-momentum:
\begin{equation}
\nabla_\mu(\rho u^\mu) = 0 ,\qquad\qquad \nabla_\mu T^{\mu\nu} \equiv \nabla_\mu (T^{\mu\nu}_\mathrm{gas} + T^{\mu\nu}_\mathrm{em}) = 0,
\end{equation}
where $\rho$ is the rest mass density, $u^\mu$ is the four-velocity of the fluid and $T^{\mu\nu}$ is the total energy-momentum tensor, with its fluid (gas) and fields (em) components, where the latter is given by
\begin{equation}
T^{\mu\nu}_\mathrm{em} = F^\mu_{\,\,\,\lambda}F^{\nu\lambda}  - \tfrac{1}{4}(F_{\lambda\kappa}F^{\lambda\kappa}) g^{\mu\nu},
\end{equation}
with $g_{\mu\nu}$ the metric tensor.
The Faraday tensor (and its dual) can be conveniently written as functions of the four-velocity and the magnetic $b^\mu$ and electric $e^\mu$ fields measured in the fluid rest frame:
\begin{equation}
\begin{array}{lcl}
F^{\mu\nu}  & = & u^\mu e^\nu - u^\nu e^\mu + \epsilon^{\mu\nu\lambda\kappa}b_\lambda u_\kappa, \\
F^{*\mu\nu} & = & u^\mu b^\nu - u^\nu b^\mu - \epsilon^{\mu\nu\lambda\kappa}e_\lambda u_\kappa,
\end{array}
\end{equation}
where $\epsilon^{\mu\nu\lambda\kappa}$ represents the Levi-Civita pseudo-tensor. It is then convenient to express the two energy-momentum tensor contributions as\footnote{ Unfortunately, in \citet{DelBuc2018,MMBZ2019} the last term with the Levi-Civita tensors was forgotten, while all the other equations in those works are correct.}:
\begin{equation}
\begin{array}{lcl}
T^{\mu\nu}_\mathrm{gas}    & = & \rho h \, u^\mu u^\nu + pg^{\mu\nu}, \\ \noalign{\medskip}
T^{\mu\nu}_\mathrm{em} & = & (e^2 + b^2)\left(u^\mu u^\nu +\DS\tfrac{1}{2}g^{\mu\nu}\right)- e^\mu e^\nu - b^\mu b^\nu + \\ \noalign{\medskip}
& & + ( u^\mu\epsilon^{\nu\lambda\kappa\rho} + u^\nu\epsilon^{\mu\lambda\kappa\rho} ) e_\lambda b_\kappa u_\rho ,
\end{array}
\end{equation}
where $h$, $\varepsilon$ and $p$ are the gas specific enthalpy, energy density and pressure, respectively, measured in the fluid rest frame and related by $\rho h = \varepsilon + p$. Notice that in the non-resistive case of ideal MHD the condition of a vanishing electric field must hold in the frame comoving with the fluid, that is $e^\mu = 0$, and the expression for $T^{\mu\nu}_\mathrm{em}$ simplifies considerably.


In numerical relativity, it is convenient to split time and space according to a so-called \textit{Eulerian} observer (e.g. \citealt{ECHO2007,REZZAN2013}). Here we assume for simplicity a flat spacetime, and this duty will be easily accomplished. The fluid four-velocity is split as:
\begin{equation}
 u^\mu = (\sqrt{1 + u^2},\vec{u}) \equiv (\gamma,\gamma\vec{v}),
\end{equation}
where $\gamma$ is the Lorentz factor and $\vec{v}$ is the three-dimensional velocity.
The relation between the rest-frame ($e^\mu$,\,$b^\mu$) and the standard laboratory (Eulerian) frame ($\vec{E}$,\,$\vec{B}$) electromagnetic field components is:
\begin{equation}
\begin{array}{lll}
e^\mu & = & (\vec{u}\cdot\vec{E}, \,\gamma \, \vec{E} + \vec{u}\times\vec{B}), \\ \noalign{\medskip}
b^\mu & = & (\vec{u}\cdot\vec{B}, \, \gamma \, \vec{B} - \vec{u}\times\vec{E}).
\end{array}
\label{emubmu}
\end{equation}

In the resistive case $e^\mu$ is not vanishing, and the simplest possible form of the (relativistic) Ohm's law assumes that it is proportional to the current density measured in the fluid rest frame $j^\mu = J^\mu + (J^\nu u_\nu) u^\mu$, that is:
\begin{equation}
e^\mu = \eta j^\mu,
\end{equation} 
where $\eta$ is the resistivity coefficient (which here, for the sake of simplicity, is assumed to be a scalar function, or even constant). In the assumed split, the explicit expression of the four-current density for flat metric becomes:
\begin{equation}
J^\mu = (q,\vec{J}) = (q \, , \, q\vec{v} + \eta^{-1}[\gamma\vec{E} + \vec{u}\times\vec{B} - (\vec{E}\cdot\vec{u})\vec{v}]),
\end{equation}
in which $q$ represents the charge density measured in the laboratory frame. 
Notice that the spatial current density $\vec{J}$ contains a standard advection term (of $q$) and the resistive term, 
proportional to $1/\eta$, that is expected to be large in high-Reynolds number astrophysical plasmas.

Splitting the time and space components of the equations themselves, the evolutionary system of resistive relativistic MHD (RRMHD from now on) is, in vectorial form:
\begin{equation}
\begin{array}{l}
\DS\pd{D}{t} + \nabla\cdot(D\vec{v}) = 0 , \\ \noalign{\medskip}
\DS\pd{\vec{m}}{t} + \nabla\cdot [\rho h \, \vec{u}\vec{u} - \vec{E}\vec{E} - \vec{B}\vec{B} + (p + u_\mathrm{em})\tens{I}] = 0 , \\ \noalign{\medskip}
\DS\pd{\cal E}{t} + \nabla\cdot\vec{m} = 0 , \\ \noalign{\medskip}
\DS\pd{\vec{B}}{t} + \nabla\times\vec{E} = 0 , \\ \noalign{\medskip}
\DS\pd{\vec{E}}{t} - \nabla\times\vec{B} = -\vec{J},
\end{array}
\end{equation}
respectively the equations for the conservation of mass, momentum, energy, and the two evolutionary Maxwell equations, with $u_\mathrm{em} = (E^2 + B^2)/2$ the electromagnetic energy density.
The conserved quantities are, respectively, the density in the laboratory frame $D$, the total momentum density $\vec{m}$, the total energy density ${\cal E}$, the magnetic $\vec{B}$ and electric $\vec{E}$ fields.
The relation from primitive ($\rho$,$\vec{v}$,$p$) to conserved fluid quantities is:
\begin{equation}
\begin{array}{lcl}
D & = & \rho\gamma,  \\  \noalign{\medskip}
\vec{m} & = & \rho h\gamma\vec{u} + \vec{E}\times\vec{B}, \\ \noalign{\medskip}
{\cal E} & = & \rho h\gamma^2 - p + u_\mathrm{em},
\end{array}
\end{equation}
while, contrary to non-relativistic MHD, an analytical relation from conserved to primitive variables is not possible.
The set of RRMHD equations is completed by the non-evolutionary Maxwell's constraints:
\begin{equation}
\nabla\cdot\vec{B} = 0 , \qquad\qquad \nabla\cdot\vec{E} = q,
\end{equation}
and by a suitable equation of state relating the thermodynamical variables, in particular providing $h=h(\rho,p)$ (see next subsection).

Notice that in the ideal MHD case, when $\eta\to 0$, the vanishing of $e^\mu$ translates 
to the condition $\vec{E} = - \vec{v}\times\vec{B}$, the same as in classical MHD.
Therefore, the field $\vec{E}$ is now a derived quantity, and the last evolution equation of the RRMHD set does not need to be solved. 
Moreover, in this case Eqs. (\ref{emubmu}) become:
\begin{equation}
\begin{array}{lll}
e^\mu & = & 0, \\ \noalign{\medskip}
b^\mu & = & (\vec{u}\cdot\vec{B}, \vec{B}/\gamma + (\vec{u}\cdot\vec{B})\,\vec{v}).
\end{array}
\end{equation}

\subsection{The Taub Equation of State (EoS)}
\label{sec::Eos}
%

The RRMHD system must be closed by an Equation of State (EoS), describing the thermodynamical 
properties of the gas through a relation between the (specific) enthalpy 
and the temperature function $\Theta = p/\rho$. In numerical relativity hydro and MHD 
simulations the constant $\Gamma-$law for an ideal gas is typically assumed:
\begin{equation}
h = 1 + \DS\frac{\Gamma}{\Gamma - 1}\Theta,
\end{equation}
where $\Gamma$ is the specific heat ratio, whose value lies between $4/3$ (ultra-relativistic plasma) 
and $5/3$ (non-relativistic plasma).
A major drawback of such approach is that the Taub's inequality \citep{Taub1948}, which is required in order 
to gain consistency with the relativistic kinetic theory :
\begin{equation}
(h - \Theta)(h - 4\Theta) \geq 1,
\end{equation}
may not be fulfilled.

From the theory of relativistic perfect gas \citep{Synge1957}, a rigorous yet time-consuming relation 
(which includes modified Bessel functions) can be recovered. An alternative approach, 
which always ensures the Taub's relation (with the equal sign), is to assume (Taub EoS from now on):
\begin{equation}
h = \DS\frac{5}{2}\Theta + \sqrt{\DS\frac{9}{4}\Theta^2 + 1}.
\end{equation}
The above EoS is an approximation of the rigorous Synge's function $h(\Theta)$ (they differ for 
few percent values), it is much faster to compute numerically, and it retains the two physically 
important limiting values. It has been applied to the ideal relativistic MHD equations by \citet{MigMck2007} and later extended by \citet{Mizuno2013} to resistive relativistic MHD applications.
All the simulations presented in this paper, if not stated otherwise, involve the above Taub EoS.

\subsection{The IMEX scheme and numerical methods}
\label{sec:IMEX}
%

Let us start by discussing the time-stepping algorithm. The set of RRMHD equations can be rewritten in a more compact form:
\begin{equation}\label{eq::RRMHD_STIFF}
\partial_t\cU = -\nabla\cdot\tens{F}(\cU) + \cS_e + \DS\frac{1}{\eta}\cS(\cU) = \cR(\cU) + \DS\frac{1}{\eta}\cS(\cU),
\end{equation}
where $\cU$ is the vector of the conserved variables, $\cR$ contains the divergence of the fluxes $\tens{F}$ 
and the standard source terms $\cS_e$, whereas the $\cS$ includes the \textit{stiff} terms, 
i.e. those proportional to $1/\eta \gg 1$ (the resistive part of the current density, 
in the equation for the electric field), requiring some special numerical treatment.
Because of the stiffness of the RRMHD equations, purely explicit algorithms would require a drastically small timestep in order to encompass the time-scales typical of the nearly ideal regime.
Several strategies have been developed in order to overcome such issue.
For instance, \citet{Komissarov2007,TakIno2011,Mizuno2013} employed a Strang splitting algorithm. 
However, as pointed in \citet{RBPetal2019}, due to the strong assumptions of such approach
(i.e. that the magnetic field and the fluid velocity remain unaltered during the evolution of the electric field),
the timestep required to achieve convergence again decreases dramatically in the regimes 
typical of astrophysical plasmas.

On the other hand, the \textit{Strong Stability Preserving} (SPP) IMplicit-EXplicit (IMEX) Runge-Kutta (RK) 
schemes by \citet{ParRus2005} have shown great stability properties without additional assumptions, 
and are frequently adopted for both special and general relativistic resistive MHD simulations 
\citep{PLRR2009,DzPLBB2016,TDzBB2020}. 
In this paper we adopt the IMEX-RK SSP3 (3,3,2) scheme to solve Eq. \ref{eq::RRMHD_STIFF}:
\begin{equation}\label{eq::IMEX}
\begin{array}{lcl}
\cU^1 & = & \cU^n + a\DS\frac{\Delta t}{\eta}\cS^1, \\ \noalign{\medskip}
\cU^2  & = & \cU^n + \Delta t\cR^1 + \DS\frac{\Delta t}{\eta}\left[\left(1 - 2a\right)\cS^1 + a\cS^2\right], \\ \noalign{\medskip}
\cU^3  & = & \cU^n + \DS\frac{\Delta t}{4}\left[\cR^1 + \cR^2\right] + \DS\frac{\Delta t}{\eta}\left[\left(\DS\frac{1}{2} - a\right)\cS^1 + a\cS^3\right], \\ \noalign{\medskip}
\cU^{n+1}  & = & \cU^n + \DS\frac{\Delta t}{6}\left[\cR^1 + \cR^2 + 4\cR^3\right] + \DS\frac{\Delta t}{6\eta}\left[\cS^1+ 
 \cS^2 + 4\cS^3\right],
\end{array}
\end{equation}
where $a = 1-1/\sqrt{2}$, which is second order accurate in time. Note that Eq. \ref{eq::IMEX} holds for any general set of hyperbolic differential equations with stiff terms, with the coefficient derived from the Butcher's tableau \citep{ParRus2005}. The extension of the IMEX scheme to a generic EoS, and to the Taub one in particular, can be found in the next sub-section.

All the simulations have been performed by using the PLUTO code \citep{PLUTO2007,PLUTO2012}, adopting a fourth-order {\it Piecewise Parabolic Method} (PPM) \citep{MPB2005,Mignone2014} for spatial reconstruction of variables at intercell faces, to compute numerical fluxes.
For the sake of simplicity and ease of reproducibility of results, we choose a {\it Harten, Lax, van Leer} (HLL) \citep{HLL1983} approximate Riemann solver, with local maximum characteristic velocities equal to the speed of light (recall that we are dealing with a hot, magnetized plasma with high flow velocities), since it proves to be very robust and no qualitative differences have been noticed by employing the more accurate estimates for the maximum eigenvalues \citep{LAAMetal2005,ECHO2007}. For similar reasons, we choose not to adopt a specific Maxwell Riemann solver for the induction equation \citep{MMBZ2019}. The use of more diffusive Riemann solvers is expected to be compensated here by performing high-resolution runs and using high-order spatial reconstruction methods \citep{DZBL2003,ECHO2007}.

As far as the methods for preserving the constraints in Maxwell equations, we adopt the divergence cleaning (GLM) method \citep{DKKMSW2002,MigTze2010} and we do not evolve the charge density, which is directly replaced by the divergence of the electric field \citep{BucDel2013,BDzB2014}.
A comparison between the divergence cleaning method adopted in this paper and the {\it Upwind Constrained Transport} (UCT) for both the magnetic \citep{LonDel2004,ECHO2007} and electric field \citep{MMBZ2019} is shown in Appendix \ref{App::EosTests}, through a Kelvin-Helmholtz instability benchmark.
These choices are again motivated by simplicity, ease of implementation and reproducibility (in different codes where not all methods may be available), given that from our benchmarks accuracy and robustness are preserved anyway.

In order to ensure stability in the more troublesome zones, we switch on the shock flattening procedures of the PLUTO code \citep{MigBod20006}. 
In presence of zones with negative pressure, we set the latter to a floor value of $p_f = 10^{-5}$ while redefining conservative variables accordingly.

\subsection{The IMEX scheme for a generic EoS}
\label{sec:IMEX-EoS}
%

Here we derive, for the first time to our knowledge, the extension of the IMEX scheme for 
any EoS, in particular for the Taub equations of state.
The implicit step of the IMEX scheme is solved by finding the roots of the equation:
\begin{equation}
\vec{f}(\vec{u}) \equiv \vec{m} - [Dh(\vec{u})\vec{u} + \vec{E}(\vec{u})\times\vec{B}] = 0,
\end{equation}
through a Newton-Broyden algorithm (see eq. 59 of \citealt{MMBZ2019}),
where $D$, $\vec{m}$, ${\cal E}$, and $\vec{B}$ are the conserved variables, known in the
conversion step to the primitive variables.
The explicit form of the Jacobian is:
\begin{equation}
{\tens J}_{ij} = \pd{f_i}{u^j} = -Dh\,\delta_{ij} - D\,u_i\pd{h}{u^j} - \epsilon_{ilm}\pd{E^l}{u^j}B^m,
\end{equation}
where $\epsilon_{ilm}$ is the 3D Levi-Civita symbol.
The electric field components and their derivatives can be easily calculated by using the prescription of \citet{TDzBB2020} (the electric field does not depend on the EoS).

Once the electric field is computed (as function of the four-velocity $\vec{u}$), 
the fluid pressure can also be derived from the energy equation.
If we define ${\cal E}_g = {\cal E} - u_\mathrm{em}$, which in turn
depends on $\vec{u}$, we can obtain $p=p(\gamma)$, hence $p=p(\vec{u})$.
In the ideal case the expression is:
\begin{equation}
  p = \frac{{\cal E}_g - D\gamma}{\gamma^2 \Gamma/(\Gamma - 1)-1},
\end{equation}
while, if the Taub EoS is employed, it can be derived by solving the quadratic equation:
\begin{equation}
p^2(1 - 5\gamma^2 + 4\gamma^4) + p(2{\cal E}_g - 5{\cal E}_g\gamma^2) + {\cal E}^2_g - D^2\gamma^2 = 0.
\end{equation}
The pressure is needed to compute the temperature function $\Theta = p/\rho = p\gamma/D$, 
hence the enthalpy $h=h(\Theta)$ is recovered, and it is possible to compute ${\vec f}({\vec u})$.
The derivatives of the specific enthalpy, to be inserted in the Jacobian $\tens{J}_{ij}$, can be computed as:
\begin{equation}
D \DS\pd{h}{u^j} = - \frac{\dot{h}}{\gamma^2\dot{h} - 1}\left[(\gamma h D + p) \frac{u_j}{\gamma} + \gamma E_i \DS\pd{E^i}{u^j}\right],
\end{equation}
independently on the EoS, where
\begin{equation}
\dot{h} = \frac{\mathrm{d}h}{\mathrm{d}\Theta} = \left\{
\begin{array}{ll}
 \DS\frac{\Gamma}{\Gamma - 1}, \qquad\qquad & \text{Ideal gas}, \\ \noalign{\medskip}
 \DS\frac{5}{2} + \DS\frac{9\Theta}{2(h - 5\Theta)}, & \text{Taub}.
\end{array}\right.
\end{equation}

The implementation of the coupling between the IMEX scheme and the EoS proposed has been tested through a set of numerical benchmarks which are shown in Appendix \ref{App::EosTests}.

From the computational point of view, a crucial point is the initial guess for the 4-velocity in the iteration scheme, which may not be trivial for more complex configurations of the RRMHD variables.
Towards the ideal regime (e.g. $\Delta t/\eta \gtrsim 10$), the ideal guess allows for convergence in less than 5 iteration.
On the other hand, in the more diffusive regimes, the ideal solution becomes brittle and therefore the solution at the previous timestep is taken.
However, when the velocity is non-negligible in any of the three components, the convergence may not be reached with one of the two guesses mentioned previously.
In such case, the Lorentz factor grows exponentially until it reaches unphysical values. To circumvent this scenario, we put a flag on the Lorentz factor.
In case the upper value of $\gamma = 1000$, which in the simulations presented in the paper is a clear sign of lack of convergence of the implicit step, we reset the primitive variables to their value at the previous timestep, while the velocity is set to zero in all the components.
By adopting this fix, the stability of the simulations has strongly improved.
As a drawback, the number of iterations in the challenging cells has increased from $\sim5$ to $\sim 15$.
However, we point out that this increase in the number of iterations occurs only in few cells and at very few steps, producing negligible additional computational time.
\section{Jet propagation and resistive models}
\label{sec:results}
%

Here we discuss the propagation of relativistic jets with finite conductivity.
At first we briefly describe the numerical setup, then we investigate the effects of the resistivity.
Consequently we introduce more consistent diffusivity models and we discuss the results obtained by employing non-constant physical resistivity.

\subsection{Numerical setup}
\label{sec:setup}
%

We consider a 2D axisymmetric setup in cylindrical coordinates ($r$, $\phi$, $z$) with $r\in[0,25 \,r_j]$, $z\in [0,50 \,r_j]$ (invariance is assumed in the toroidal direction $\phi$), and uniform grid resolution of  $1200\times2400$ computational zones. The magnetized relativistic jet is injected into the domain from the lower boundary ($z = 0$) in the region $r < r_j$, where $r_j = 1$ is the characteristic length used for normalization, which corresponds to 48 grid cells in both $r$ and $z$ directions.

The jet is initialized with uniform density $\rho_j = 1$ and vertical velocity $\vec{v}_j = (0,0,v_j)$, specified by the Lorentz factor $\gamma_j = 10$ (i.e. $v_j \simeq 0.99$).
Its magnetic field structure is the same as in \citet{MUB2009}.
It consists of a variable toroidal component, defined as:
\begin{equation}
B^\phi(r) = \gamma_j b_m \,\mathrm{min}\left(\DS\frac{r}{r_m},\DS\frac{r_m}{r}\right),
\qquad 0\leq r\leq r_j,
\end{equation}
where $r_m = r_j/2$ is the radius where the toroidal field attains its maximum value $\gamma_j b_m$. This is determined by:
\begin{equation}
b_m = \DS\sqrt{\DS\frac{4p_j\sigma_\phi}{(r_m/r_j)^2[1 - 4\log (r_m/r_j) - 2\sigma_\phi]}}, 
\end{equation}
where we set $\sigma_\phi = 0.3$. 
The value of the pressure $p_j$ is retrieved by imposing the Mach number:
\begin{equation}
 M = \frac{v_j}{c_s} = v_j\DS\sqrt{\DS\frac{\rho_j}{\Gamma p_j} + \DS\frac{1}{\Gamma - 1}},
\end{equation}
and here we choose $M=6$ and $\Gamma = 5/3$, independently on the EoS employed.
The thermal pressure distribution inside the jet can then be recovered assuming radial momentum equilibrium, and we find:
\begin{equation}
p(r) = p_j + b_m^2\left[1 - \mathrm{min}\left(\DS\frac{r^2}{r_m^2},1\right)\right], \qquad 0\leq r\leq r_j,
\end{equation}
hence the pressure has a maximum at the axis for $r=0$, reaches its minimum value $p_j$ for $r=r_m$,
and remains constant beyond it.
The poloidal magnetic field is instead purely vertical and constant everywhere:
\begin{equation}
B^z = \DS\sqrt{\sigma_z[(r_m/r_j)^2 b_m^2 + 2p_j]},
\end{equation}
with $\sigma_z=0.7$. 
The two free \textit{magnetization parameters}
$\sigma_\phi$ and $\sigma_z$
are actually defined through radial averages of quantities like $2b^2/p$, see
\citet{MUB2009} for their actual definitions.
The electric field, unless stated otherwise, 
is initialized inside the jet to its ideal value $\vec{E} = -\vec{v}\times\vec{B}$,
providing a purely radial component $E^r = v_j B^\phi(r)$.

The external ambient medium is static and uniform, with density $\rho_a = 10^3 \rho_j$, gas pressure $p_j$, purely vertical magnetic field $B^z$ (as in \citealt{LAAMetal2005,BecSto2011}), vanishing electric field. 
Hence, at the jet boundary $r=r_j$, the quantities $\rho$, $v^z$, $B^\phi$, and $E^r$ are discontinuous.

All the simulations are run until $t = 300$ (time will be expressed from now on in units $r_j/c$).
In order to easily distinguish the jet plasma from that coming from the ambient medium, 
we follow the evolution of 
an additional equation describing the advection of a tracer $f$, 
which vanishes in the ambient and is injected with the value $f=1$ in the jet region.

\subsection{Constant resistivity models}
\label{sec:etaregime}
%

\begin{figure*}
    \centering
    \includegraphics[width=0.99\textwidth]{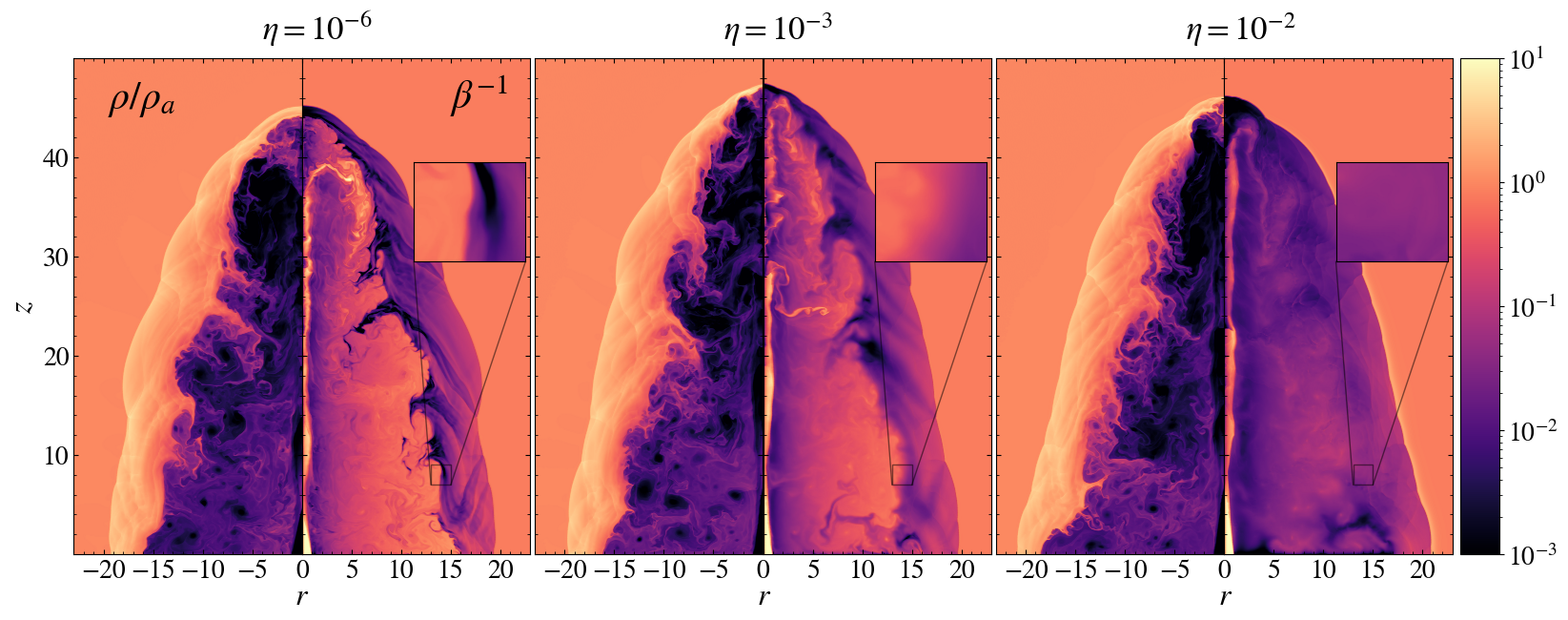}
    \caption{Density (left panels) and inverse of the plasma-$\beta$ (right panels) for different values of the resistivity, in logarithmic scale at $t = 300$. The left, middle and right panel blocks show results computed with resistivity values of $\eta = 10^{-6}$, $\eta = 10^{-3}$ and $\eta = 10^{-2}$ respectively.}
    \label{fig::rho_etacons}
\end{figure*}

In order to investigate the impact of the resistivity on the jet propagation process, we performed three simulations with constant (in space and time) resistivity (low, $\eta = 10^{-6}$, medium, $\eta = 10^{-3}$, and high, $\eta = 10^{-2}$).
We also ran a simulation (not shown here for the sake of simplicity) with the same numerical setup, 
numerical schemes and infinite conductivity as an additional validation.
Differences between the ideal simulation and the simulation with the lowest resistivity value 
$\eta = 10^{-6}$ showed no qualitative difference. Therefore we expect the numerical diffusivity to lie between $10^{-6}$ and $10^{-3}$, although we chose a lower value of eta as proof of code robustness.

In Fig. \ref{fig::rho_etacons} we show maps, at the final time and for the three reference 
resistive simulations, of the rest mass density and of the inverse of the plasma$-\beta$,
that is $\beta^{-1}=B^2/2p$, the ratio of magnetic to thermal pressure.
Together with the proper relativistic magnetization (defined as $\sigma = B^2/\rho h$, or
$\sigma = b^2/\rho h$), the above classical definition of the inverse plasma-$\beta$ 
is very often used as a proxy for magnetization in studies of the reconnection process 
in the non-relativistic and in the relativistic regime \citep{DzPLBB2016,RPSK2019,PMB2021}.
As a preliminary consideration we notice that the level of refinement (determined by 
the turbulent structures shown in the density and the plasma-$\beta$ maps) of the low resistivity simulation, is comparable with the one obtained by (\citealt{MUB2009}, see also Appendix \ref{App::turbulence}), who employed the less diffusive HLLD Riemann solver, linear reconstruction and a $3^{\rm rd}$-order time integration using a Runge–Kutta algorithm \citep{GST2001}.
More accurate and less diffusive Riemann solvers \citep{MAAR2018,MMBZ2019} would be likely to suppress the numerical diffusivity and enhance even more the turbulent jet structure (when looking at the fluid variables); nevertheless, in presence of physical resistivity, the spatial scale of the magnetic turbulence should not be dictated by the numerical methods (see the resolution study of Appendix \ref{App::turbulence}).
However, we confirm here that for relativistic MHD the HLL approximate Riemann solver combined to high-order spatial reconstruction \citep{DZBL2003,ECHO2007} is a robust and efficient compromise also for resistive simulations.

\begin{figure}
    \centering
    \includegraphics[width=0.45\textwidth]{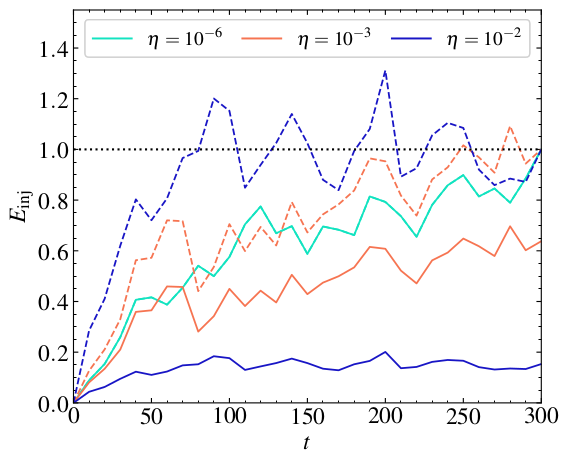}
    \caption{Jet injected electromagnetic energy. The solid lines represent the energy normalized to the final value of the run with $\eta = 10^{-6}$, while the dashed lines represent the same quantities normalized to the final values of each corresponding simulation.}
    \label{fig::energy_etacons}
\end{figure}

\begin{figure}
    \centering
    \includegraphics[width=0.45\textwidth]{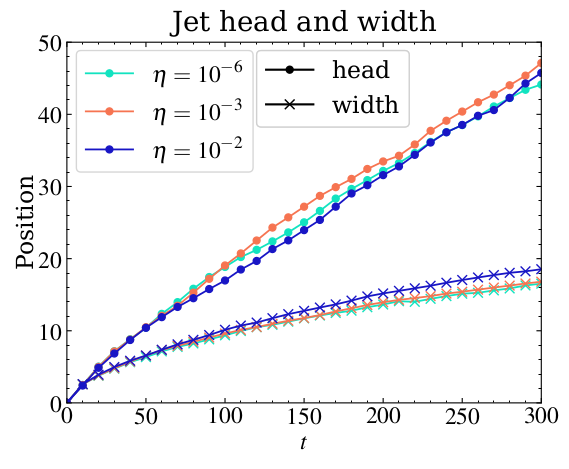}
    \caption{Jet head (circles) and width (crosses) positions for different values of the resistivity computed as functions of time.}
    \label{fig::jethead_etacons}
\end{figure}

\begin{figure*}
    \centering
    \includegraphics[width=0.99\textwidth]{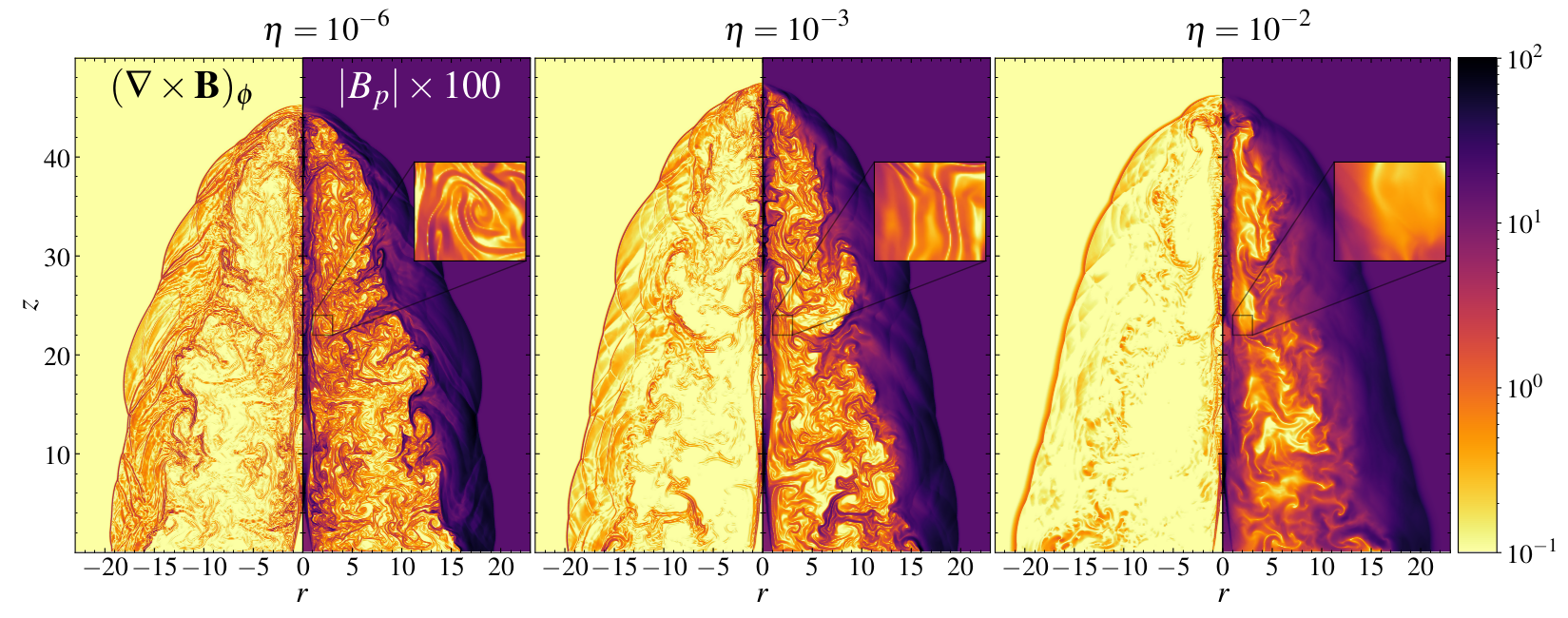}
    \caption{Toroidal component of the magnetic field's curl (left panel blocks, representing a proxy for the toroidal current) and poloidal magnetic field (right panel blocks) for different values of the resistivity, in logarithmic scale at $t = 300$. The left, middle and right panel blocks show results computed with resistivity values of $\eta = 10^{-6}$, $\eta = 10^{-3}$ and $\eta = 10^{-2}$ respectively.}
    \label{fig::jphi_etacons}
\end{figure*}

The most prominent differences are related to the electromagnetic field, which is strongly affected by the physical resistivity present in our simulations.
We notice (see the zoomed panels of Fig. \ref{fig::rho_etacons}) a much steeper gradient in the magnetization at the interface (contact discontinuity) between the jet plasma and the shocked ambient medium (from 2 orders of magnitude in the low-$\eta$ case to a smooth transition in the high-$\eta$ case).
Such a feature is not visible while looking at the density profile, and it is related to the suppression, due to the smearing of magnetic gradients, of the magnetic field turbulence caused by the interaction between the jet gas and the ambient medium.
As a result, the jet magnetization is suppressed for the largest value of $\eta$.

For low and moderate values of the resistivity (i.e. $\eta \lesssim10^{-3}$), the jets show multiple zones with an average magnetization between 0.1 and 10, 
which is comparable to previous RMHD reconnection studies \citep{DzPLBB2016,RPSK2019}, suggesting that large-scale magnetic reconnection and plasmoids formation may occur in such locations (see the next sub-section).
Note that in the cited works the plasma magnetization was set as an initial parameter, which is independent on the resistivity; here, despite the same initial physical conditions
both at injection and in the ambient plasma for all our simulations, the final jet magnetization is strictly interrelated with the physical resistivity.

The impact of the resistivity reflects on the large-scale evolution of the field, as it can be shown by computing the injected electromagnetic energy, defined as (see \citealt{PMQ2019})
\begin{equation}
E_{\rm inj} = \int u_\mathrm{em}\,f\, 2\pi r \,\mathrm{d}r\,\mathrm{d}z,
\end{equation}
where $f$ is the jet tracer. This is shown in Fig \ref{fig::energy_etacons} as a function of time.
The solid line represent the values of the injected energy normalized to the final value of the run with $\eta = 10^{-6}$.
In the very early evolutionary phases, the middle and low resistivity cases are almost indistinguishable.
However, around $t\sim35$, the two cases start showing some differences since the jet head has already started interacting with the surrounding medium, slowing down the propagation and triggering the turbulence.
Such difference is much more prominent while looking at the high diffusivity cases, in terms of both time scale and suppression of the electromagnetic energy.
At time $t\sim 300$ the ratio between the injected energy is $\sim0.64$ between the middle and the low resistivity and $\sim0.15$ between the high and the low resistivity cases.

The dashed lines of Fig. \ref{fig::energy_etacons} represent the fraction of injected energy normalized respect to the final value of each simulation (the cyan solid and dashed lines overlap since their normalization is equivalent).
We notice that the cases with higher resistivity reach a higher fraction of their final injected energy at much earlier times.
Such differences can be also related to the contribution provided by the resistive field, 
which becomes less negligible for higher values of $\eta$ (see a wider discussion in \ref{sec:res_efield}).
This, in addition to the higher field diffusion (as shown by the maps of $\beta^{-1}$), may contribute to explain this behavior.

On the other hand, the impact of the lower jet magnetization and higher resistive electric field for increasing values of $\eta$ on the jet shape is much less evident (see Fig. \ref{fig::jethead_etacons}).
This is shown by the low and medium resistivity cases whose jet width (defined as the largest $r-$coordinate of the interface between the jet and the cocoon) differs from each other by $\lesssim10\%$.
On the other hand, the high resistivity case yields a slower (for most of the evolution) and wider jet. 
The impact of the resistivity on the jet shape becomes more relevant after the turbulence has fully developed within the jet. 
In particular, differences in the jet width or head (defined as the largest $z-$coordinate of the interface between the jet and the cocoon)
are not seen ($\lesssim1\%$) until $t\gtrsim70$.

\subsubsection{The formation of current sheets}
\label{sec:Magfieldrev}
%

\begin{figure*}
    \centering
    \includegraphics[width=0.9\textwidth]{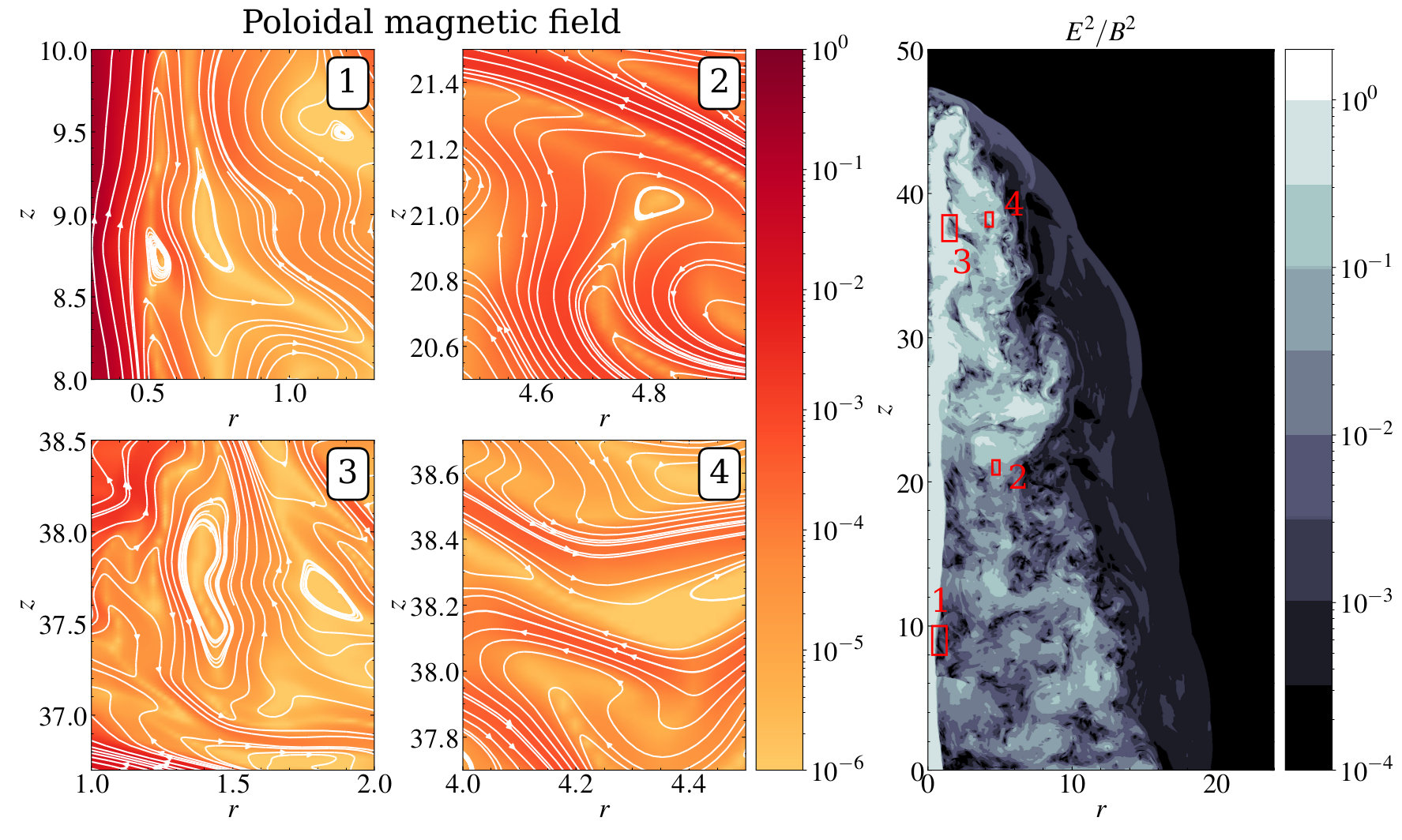}
    \caption{Current sheet forming in the turbulent jet region for the case $\eta = 10^{-3}$ at $t = 300$. The right panel shows the ratio between the electric and magnetic components of the electromagnetic energy. The 4 left panels show zoomed portions of the jet domain (see the rectangles in the right plot), where the poloidal magnetic field (color) and magnetic field lines (white lines) are plotted in order to highlight the reversal of the magnetic field and the formation of magnetic islands due to magnetic reconnection.}
    \label{fig::sheet_etacons}
\end{figure*}

As we have seen, the most prominent differences among the propagating jets at various values
of (constant) $\eta$ are related to the electromagnetic field, which is strongly affected 
by its diffusion caused by the non-ideal processes occurring in our simulations.
More specifically, the higher diffusion of the magnetic field leads to a smearing 
of the poloidal component, resulting in the suppression of the flow and magnetic vortexes
typical of MHD turbulence (see the zoomed panels of Fig. \ref{fig::jphi_etacons}).

Turbulence in plasmas is known to lead to the formation of current sheets and to reconnection events,
and this is particularly important in relativistic magnetically-dominated plasmas 
where the release of energy could be huge. 
Relativistic magnetic reconnection is also becoming a paradigm for
high-energy particle acceleration (e.g. \citealt{SirSpi2014,SPG2015,PSSG2019,Medetal2021}).
The large (temporal and spatial) scale separation between plasma dissipative processes (at the electron/ion skin depth level) and jet dynamics (up to kpc scales) prevents current numerical models from resolving all the relevant physical phenomena at once (as stated in \citealt{Medetal2023}).
Therefore, we focus here on the largest scale scenario dictated by the propagation of astrophysical jets,
treated in the fluid/MHD approximation, and we cannot investigate the detailed reconnection
scales and (kinetic) physics involved at the same time.

In order to more sistematically detect possible reconnection sites, we focus on the structure of the poloidal magnetic field and its curl, that is the non-relativistic current $J_\phi = (\nabla\times\vec{B})_\phi$, as often done in the literature, even by devising specific detectors based on this quantity \cite[e.g.][]{ZUPB2013,NCMB2023}.
Therefore, in Fig. \ref{fig::jphi_etacons} we show maps of the non-relativistic toroidal current, and 
of the poloidal field from which the former quantity has been computed.
As a first consideration, we point out that current sheets and magnetic islands can also appear in simulations without an explicit physical resistivity (ideal MHD), due to the intrinsic and unavoidable diffusivity of the numerical schemes. 
Properties such as the width of the current sheets arising from the ideal simulations are strictly related to the grid resolution and the numerical schemes.
Here we have verified that our ideal runs are very close to the case with $\eta=10^{-6}$, meaning that high-order reconstruction is able to maintain at the minimum level the numerical diffusivity (comparable with the ideal simulations), in spite of the use of the HLL approximate Riemann solver.

Because of the lack of turbulence caused by the physical resistivity in the 
most dissipative case with $\eta = 10^{-2}$, the formation of current sheets is largely
reduced compared with the other two cases.
In particular, thicker current sheets are formed only in selected regions, e.g. near the axis, 
the jet head and between the jet material and the bow shock.
Conversely, a lower resistivity leads to more ubiquitous and thinner current sheet regions.
The strong dependence of the formation of current sheets with the higher magnetization, 
maintained only in the lower resistivity runs, is promising, given that astrophysical plasmas
are expected to be quasi-ideal and favourable for the turbulent reconnection scenario.

In Figure \ref{fig::sheet_etacons} we have isolated four different regions from the case $\eta = 10^{-3}$ where the 
magnetic field shows an inversion of its polarity and becomes prone to the formation 
of relativistic current sheet (which are a primary ingredient for magnetic reconnection).
Each region shows one or multiple inversions of the magnetic field lines due to the 
turbulence within the jet, and the consequent tearing instability leading to the plasmoids.
Despite the lack of resolution required in order to properly resolve the complex 
internal structure of the magnetic islands, we are able to select several portions 
of each region where the magnetic field polarity is suitable for magnetic reconnection 
and plasmoid instability, allowing us to properly detect the X- and O-points in 
current sheets (as shown in the zoomed panels).
All the panels selected (but the top right, number 2) show an average magnetization 
between 0.1 and 10 (see Fig. \ref{fig::rho_etacons}), since a lower magnetization 
may not lead to the formation of plasmoids \citep{RPSK2019}.
Moreover, as shown in the right panel of Fig. \ref{fig::sheet_etacons}, notice that
the electric and magnetic contributions to the electromagnetic density are comparable
in the jet region, mostly due to the high speed flows.
Higher resolution simulations are expected to resolve the X- and O-points to a 
better accuracy, showing the regions inside current sheets where the electric field 
becomes stronger than the magnetic field.

Despite the strong jet magnetization in terms of magnetic energy, comparable
to the thermal one (at least in the lower resistivity cases), when the interaction 
with the ambient medium starts, the resulting relativistic magnetization is strongly suppressed 
by the mixing, given that the ambient medium is a thousand times denser than the jet.
The interface at the contact discontinuity is prone to the Kelvin-Helmholtz instability,
due to the strong shear velocities, and this is the main trigger for this mixing.
However, the same instability enhances the turbulence level and favours in turn magnetic
reconnection, as shown by the presence of the X-points and magnetic islands in the zoomed panels
also present in these regions.

\subsubsection{The resistive electric field}
\label{sec:res_efield}
%

\begin{figure*}
    \centering
    \includegraphics[width=0.99\textwidth]{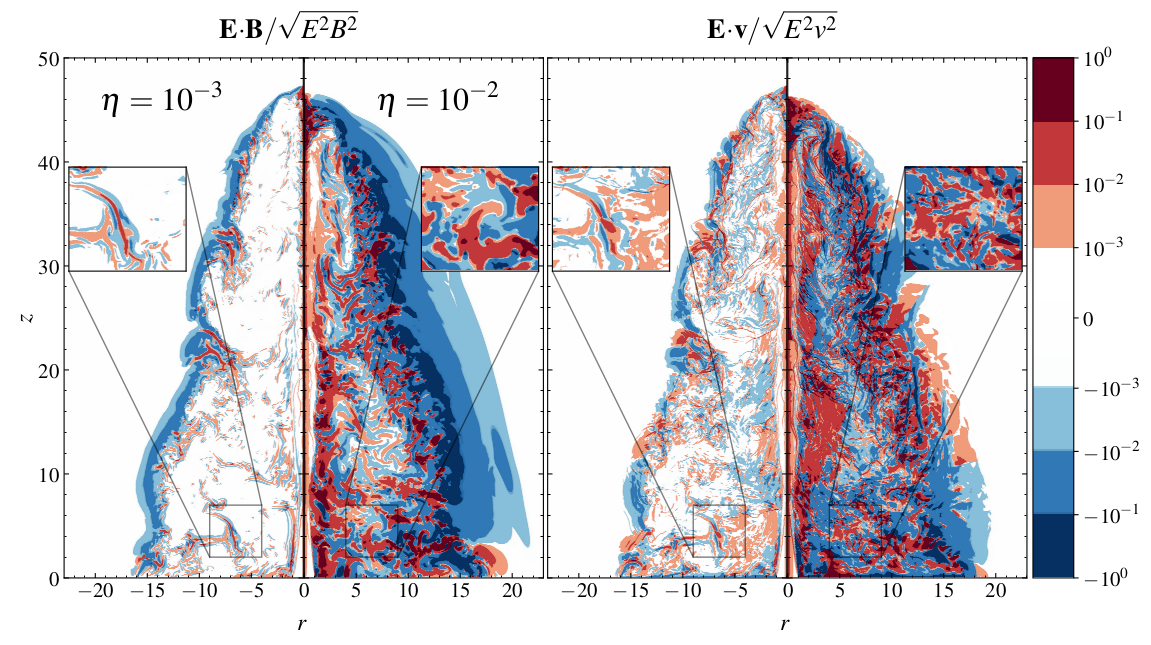}
    \caption{Normalized parallel component of the electric field. For each panel group, ${\vec E}\cdot{\vec B}$ (left panel blocks) and  ${\vec E}\cdot {\vec v}$ (right panel blocks) are shown for different values of the diffusivity at $t = 300$ in symmetric logarithmic scale. 
    The parallel components are normalised respect to the corresponding quantities. The left and right panels of each block show results computed with resistivity values of $\eta = 10^{-3}$ and $\eta = 10^{-2}$ respectively.}
    \label{fig::ebpar_etaconst}
\end{figure*}

In addition to the formation of current sheets and plasmoids, the resistive component of the electric field is also an indicator for sites where non-ideal processes may take place.
Since this term can only appear if a physical explicit
resistivity coefficient is present, magnetic reconnection in relativistic plasmas
should be addressed through consistent numerical recipes, i.e. employing resistive RMHD
schemes with explicit resistivity, as properly achieved here.

The presence of such non-ideal component of the electric field can be highlighted by
displaying quantities like ${\vec E}\cdot {\vec v}$ and  ${\vec E}\cdot {\vec B}$, since they both vanish in ideal MHD (and RMHD).
These are shown in Fig. \ref{fig::ebpar_etaconst}, the parallel electric field component with
respect to the velocity in the left panel block and the parallel electric field component with respect to the magnetic field in the right panel block, for the intermediate (left panels) and high (right panels) diffusivity models.
Notice that those components vanish at the injection region by construction (we recall that an ideal electric field is assumed there, in the radial direction). 
In presence of a high resistivity, the resistive electric field starts to arise closer to the jet base (see Fig. \ref{fig::ebpar_etaconst}).

Since the numerical resistivity is dominant in the lowest resistivity model, we do not show here the corresponding resistive electric field.
Instead, we focus on the models whose physical diffusivity overcomes the numerical dissipation, since the resistive properties can be derived more consistently.

Because of the very turbulent structure of the magnetic field, the intermediate diffusivity case yields small layers with different polarity (see the zoomed panels of Figure \ref{fig::ebpar_etaconst}).
Due to the "poor" resolution, such layers are separated by a very limited number of computational cells.
The most evident difference between the medium and the high resistivity cases is given by the the electric field orientation.
As shown by the left and middle panel blocks of Fig. \ref{fig::ebpar_etaconst}, the parallel component in the $\eta = 10^{-3}$ case shows, throughout the whole jet, an angle (given by the normalization of the scalar product between electric field and magnetic/velocity field) closer to $90\degree$ than the high resistivity case.

We also notice that in both resistive regimes the resulting spatial structure of ${\vec E}\cdot{\vec B}$ is quite smoother with respect to ${\vec E}\cdot{\vec v}$, especially for the $\eta=10^{-2}$ case where larger scales are clearly produced.
As discussed previously, this is not surprising, since the dissipation scale of the magnetic field is determined by the values of the physical resistivity, while the turbulent velocity structure is more affected by the intrinsic diffusion of the numerical algorithms.
Note that a stronger relevance of the parallel component does not necessarily mean a stronger resistive electric field, since the electromagnetic energy depends on the diffusivity as well.

In addition to the different relevance of the parallel component, the case with high resistivity shows a smoother transition of the parallel component from the jet to the outer ambient medium (where it is absent by construction).
In particular, the parallel electric field, in presence of higher resistivity, is diffused outwards more rapidly also in the region between the turbulent material and the bow shock (especially when looking at the parallel component with respect to the magnetic field).

\begin{figure}
    \centering
    \includegraphics[width=0.49\textwidth]{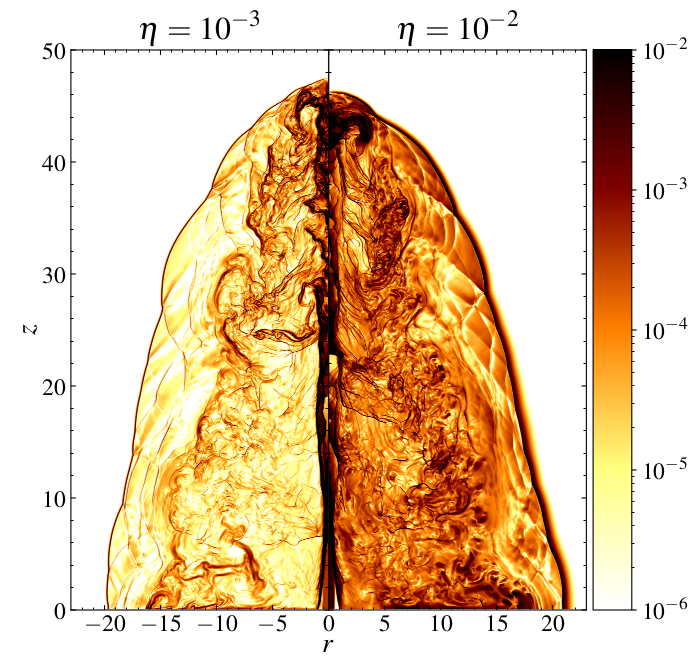}
    \caption{Jet dissipated power for the $\eta = 10^{-3}$ (left panel) and the $\eta = 10^{-2}$ (right panel) resistivity models at $t = 300$ in logarithmic scale.}
    \label{fig::power_etaconst}
\end{figure}

A final consideration can be done on the impact of the finite conductivity on the jet dissipated power, whose definition in the fluid comoving frame is:
\begin{equation}
{\cal P} = j_\mu e^\mu = e^2/\eta.
\end{equation}
Non-ideal MHD processes (e.g. resistivity) usually involve an energy conversion; in this case the magnetic energy is converted into thermal energy.
As shown in Fig. \ref{fig::power_etaconst}, a higher resistivity leads to a more efficient energy conversion rate and heat generation through all the jet up to the bow shock.
On the other hand, the case $\eta = 10^{-3}$ shows significant power dissipated near the jet head and, more generally, in the jet regions where magnetic reconnection is more likely to take place.
Such behavior is consistent with the fact that the reconnection process involves a conversion from magnetic energy to kinetic, thermal and particles (on much smaller scales than the one typical of these simulations) energy.

\subsection{Variable resistivity models}
\label{sec:variable_eta}
%

\begin{figure*}
    \centering
    \includegraphics[width=0.99\textwidth]{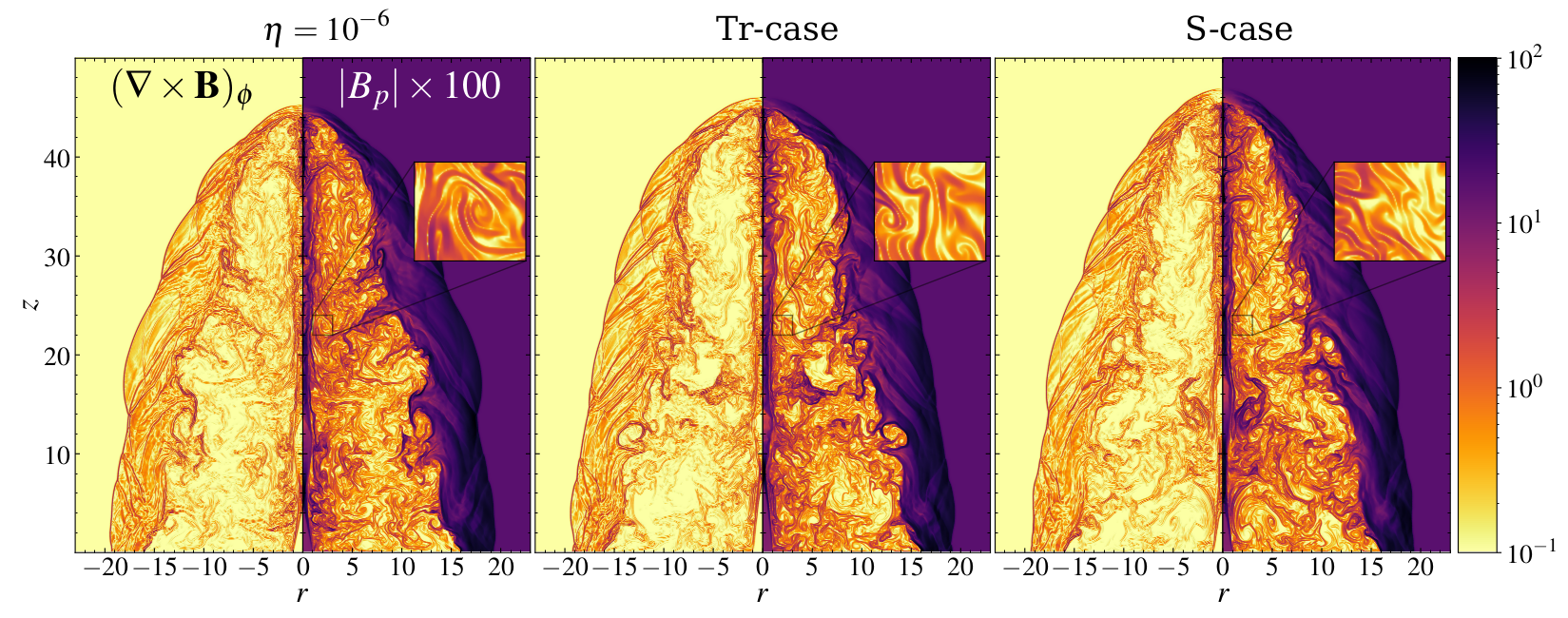}
    \caption{Toroidal current (left panel blocks, computed from the curl of the magnetic field) and poloidal magnetic field (right panel blocks) for different values of the resistivity, in logarithmic scale. The left, middle and right panel blocks show results computed respectively with a constant resistivity $\eta = 10^{-6}$, the Tr-case and the S-case.}
    \label{fig::jphi_etatr}
\end{figure*}

A constant resistivity profile, albeit a very simple and effective strategy, may not consistently model the differences between the jet and the ambient medium.
In this section we compare two more consistent, non-constant in both time and space, resistivity models.
The assumption of a resistivity confined in the jet is justified, as in \citet{FenCem2002}, by the turbulent (and thus resistive) nature of the accretion disks.
For instance, \citet{QFV2018,VFQN2019} assumed a resistivity caused by the thin accretion disk rotation which triggers the magneto-rotational instability.
A similar assumption is made by \citet{BDzB2014,TDzBB2020} in the context of mean-field dynamo in thick accretion tori.
Moreover, turbulence is also supposed to be generated in the vicinity of black holes \citep{NMPFR2022}, leading to the formation of plasmoids for up to $\sim40$ gravitational radii.
It is therefore natural to expect that the turbulent/resistive nature of the accretion flow is present also in the outflow at greater distance from the accreted matter.

A simple yet effective way to model a more consistent diffusivity profile is to relate it to the passive scalar tracer:
\begin{equation}
\eta = \max\left(10^{-6},10^{-3}f\right)
\end{equation}
Since the passive tracer has values between 0 (ambient) and 1 (jet), the resistivity will always be bound between $10^{-6}$ and $10^{-3}$.
Here the resistivity is mostly determined by the mixing of jet and ambient medium, while the electromagnetic components are only included from the full evolution of the RRMHD equations.

A more consistent resistivity profile is computed, similar to \citet{FenCem2002}, by fixing the Lundquist number $S$, yielding:
\begin{equation}
\label{eq::lundquist}
\eta = \DS\frac{v_AL}{S}
\end{equation}
where $v_A = |\vec{B}|/\sqrt{\rho h + |\vec{B}|^2}$ is the Alfv\'en speed and $L$, a typical length scale of the system, is set to be equal to the injection radius $r_j=1$.
Here, in contrast to \citet{FenCem2002}, we consider the whole magnetic field (i.e. toroidal + poloidal), instead of only the poloidal field.
We choose a value of $S = 10^3$ in order to reproduce a proper physical resistivity (i.e. higher than the numerical dissipation) in a suitable range to model astrophysical plasmas.

\begin{figure}
    \centering
    \includegraphics[width=0.45\textwidth]{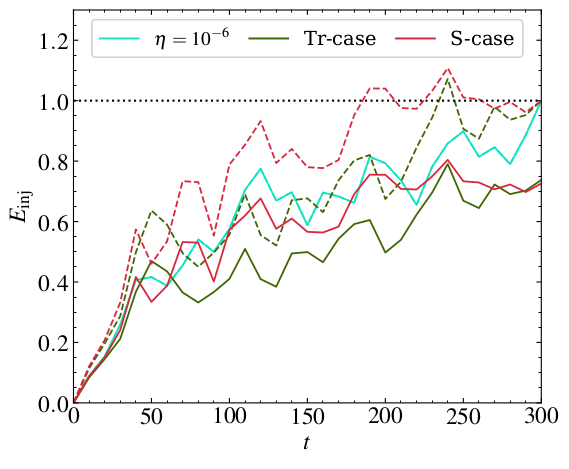}
    \caption{Jet injected electromagnetic energy for different resistivity models at $t = 300$ in logarithmic scale. The solid lines represent the energy normalized to the final value of the run with constant (in space and time) resistivity $\eta = 10^{-6}$, while the dashed lines represent the same quantities normalized to the final values of each corresponding simulation.}
    \label{fig::energy_etatr}
\end{figure}

\subsubsection{Magnetic field and magnetization}
\label{sec:variable_eta_B}
%

In Fig. \ref{fig::jphi_etatr}, we show the toroidal current and the poloidal magnetic field of the constant low resistivity model (here used as a proxy for the numerical dissipation) compared with the two more consistent resistivity profiles (henceforth Tr-case and S-case).
As a first consideration, we notice how the turbulent pattern of the poloidal magnetic field features the same level of turbulence (see Appendix \ref{App::turbulence}).
Another similarity between all these runs is the formation of current sheets, highlighted by the toroidal component of the (classical) current $(\nabla\times{\vec B})_\phi$, through the entire jet domain.
The jet head and axis, as well as the region between the jet and the bow shock, remain the locations where current sheets are more likely to be located, but the turbulent region (especially in the S-case) is equally promising as source of non-thermal particles due to magnetic reconnection.

However, the injected energy (shown in Fig. \ref{fig::energy_etatr}), in both cases with non-uniform resistivity, saturates at lower values ($\sim 75\%$ of the low constant resistivity case).
This different saturation is caused by the different values that the resistivity can assume during the temporal evolution.
By computing the average value of the jet resistivity as 
\begin{equation}
\label{eq::jet_eta}
\overline{\eta} = \DS\frac{\int \eta\,f\, 2\pi r \,\mathrm{d}r\,\mathrm{d}z}{\int f\, 2\pi r \,\mathrm{d}r\,\mathrm{d}z},
\end{equation}
we notice a different saturation of such value.
More specifically, already at $t\sim30$, the average jet resistivity has reached a value of $2.5\times10^{-4}$ for the S-case, which remains quite constant through the rest of the simulation.
On the other hand, the average jet resistivity in the Tr-case starts with a value of $\sim10^{-3}$ as the jet is injected ($t\gtrsim0$), since the jet and ambient matter have not started to interact yet; around $t\sim100$, the jet and the ambient medium have already interacted and the average jet resistivity slowly converges to $4.5\times10^{-4}$ until the end of the simulation.

This confirms the trend seen in Fig. \ref{fig::energy_etacons}, where a higher resistivity lead to a weaker injected energy.
Moreover, it explains why the case $S = 10^3$ reaches the saturation of the injected energy at earlier time; since the resistivity decreases because of the mixing, the saturation value increases as the turbulence develops.

\begin{figure*}
    \centering
    \includegraphics[width=0.99\textwidth]{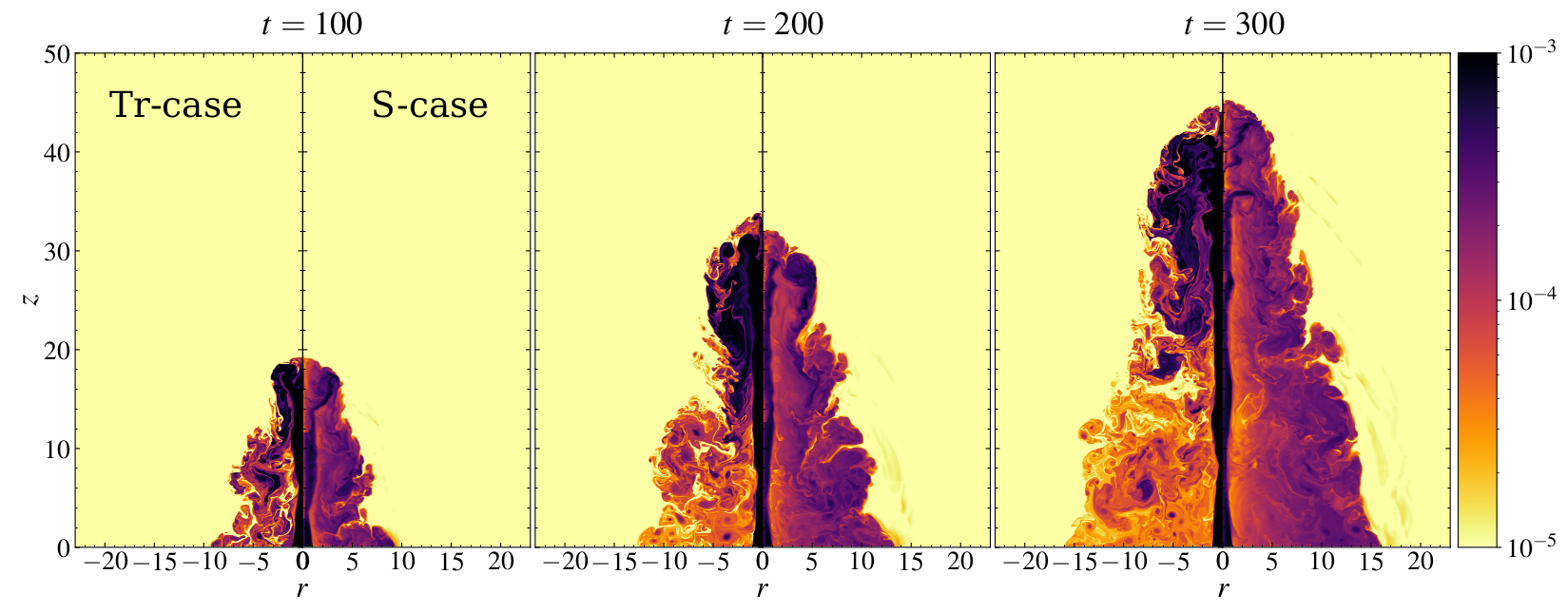}
    \caption{Resistivity computed at different times ($t = 100,200,300$ respectively in the left, middle and right panel blocks) with the two different variable resistivity model (respectively, Tr-case on the left panels and S-case on the right panels).}
    \label{fig::eta_nonconstant}
\end{figure*}

We also notice that the spatial resistivity profile shows clear differences between the different resistivity models.
More specifically, the Tr-case shows a resistivity dominated by the jet at earlier time (see the left panel of Fig. \ref{fig::eta_nonconstant}), and even at later times a difference of almost 2 orders of magnitude are present between the resistivity at the lower regions and the jet head.
On the other hand, the S-case shows a more uniform diffusivity distribution, since a stronger magnetic field triggers a higher resistivity which diffuses the magnetic field.
This "cycle" of magnetic field and diffusivity has already converged at $t\lesssim50$, as shown in the left panels of Fig. \ref{fig::eta_nonconstant}.

\subsubsection{The resistive electric field}
\label{sec:variable_eta_E}
%

\begin{figure}
    \centering
    \includegraphics[width=0.49\textwidth]{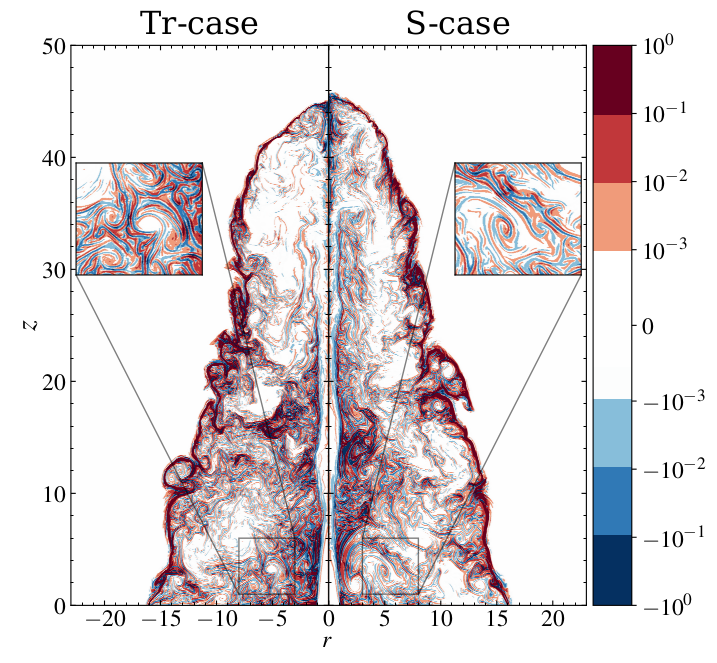}
    \caption{Parallel (normalized, as in Fig. \ref{fig::ebpar_etaconst} component of the electric field respect to the magnetic field for the Tr-case (left panel) and the S-case (right panel) resistivity models at $t = 300$ in symmetric logarithmic scale.}
    \label{fig::eb_nonconstant}
\end{figure}

As for the constant diffusivity models, the non-ideal electric field is a key component in order to investigate the role of the resistive features.
In particular, in Fig. \ref{fig::eb_nonconstant} we show the parallel component of the electric field with respect to the the magnetic field.

In both cases, the polarity of the parallel electric field near the boundary between jet and environment is the opposite with respect to the cases shown in Section \ref{sec:res_efield} (i.e. $\eta = 10^{-3}$ and $\eta = 10^{-2}$).
This feature is not surprising, since in while investigating constant diffusivity models (see Section \ref{sec:res_efield}) we noticed such inversion in presence of a numerical resistivity (which is the case at the interface between jet and ambient medium for both variable diffusivity models).

However, the parallel electric field arises near the jet base, because of the higher physical resistivity at the jet base.
The development of a strong parallel component in such region has been seen also in the $\eta = 10^{-3}$ and $\eta = 10^{-2}$ cases
As a result, the jet location where the parallel field becomes more relevant is near the axes, in agreement with the $\eta = 10^{-3}$ case.

We notice that the spatial structure of ${\vec E}\cdot{\vec B}$ is significant less smooth than the cases with constant resistivity because of the lower values of the resistivity outside the very inner propagation region.
Moreover, the two models show a consistent difference in terms of spatial distribution of the parallel field.
In particular, both cases show a relevant parallel component in the turbulent region, which is suppressed in the $\eta = 10^{-3}$ case.

The sharp gradient of the parallel field at the interface between jet and ambient is likely due to the numerical diffusivity, as in both models the value of $\eta$ drops abruptly in the jet's environment.
For the same reason, there is also a lack of diffused parallel field beyond that interface, as opposed to the models with constant resistivity (see the first panel of Fig. \ref{fig::ebpar_etaconst}). 

\section{Discussion and conclusions}
\label{sec::Summary}

In this paper we presented the first systematic numerical study, in the regime of resistive relativistic magnetohydrodynamics (RRMHD), of astrophysical relativistic jets endowed with both poloidal and toroidal magnetic field components propagating through a uniform magnetized medium. 
We combined a high resolution grid in cylindrical coordinates with high-order numerical algorithms and realistic physical assumptions in order to achieve the best level of refinement, while preserving the robustness required to overcome the numerical issues often arising when simulating relativistic flows in a strongly magnetized regime.

The following summarizes our approaches and results:

\begin{enumerate}
    \item we extended the IMEX scheme to the use of the Taub equation of state, in order to couple an accurate and robust evolution scheme for the electric field in non-ideal conditions and the proper treatment of the gas varying conditions, from the jet to the ambient.
    We have also augmented the stability of the IMEX scheme by improving the choice of the initial primitive variables during the implicit step;

    \item we have initially investigated the role of a constant (in space and time) resistivity coefficient. 
    More specifically, we found that a high resistivity coefficient, say $\eta\sim 10^{-2}$ in typical code units, corresponding to roughly a Lundquist number of $S\sim 100$ (normalized to the injection length and for an Alfvén speed close to the speed of light), suppresses almost completely the turbulence in the propagating jet and therefore the formation of relativistic current sheets. 
    Conversely, a lower resistivity (at least $\eta \lesssim 10^{-3}$, or $S\gtrsim10^3$) triggers the formation of multiple zones which could be prone (in terms of magnetization and field topology) to the plasmoid instability;

    \item the electric field, in presence of a physical resistivity, develops a component parallel to the magnetic field which can be used as an indicator of non-ideal processes occurring at the turbuent scale, even in absence of magnetic reconnection events.
    We have found that different values of the resistivity strongly impact the arising  of the resistive/parallel electric field, leading to a different spatial structure and dissipated power;

    \item the combination of high-order numerical algorithms and high resolution enables us to find with increased accuracy the formation of X- and O-points within the jets, despite the lack of resolution required to resolve the particles acceleration scales.
    However, selected zones of the large scale jets can provide more consistent recipes in order to investigate the reconnection process at smaller scales;
    
    \item we have also adopted and compared more consistent diffusivity models with a variable coefficient $\eta$, one based on a passively evolved  scalar jet tracer, and another one fixing a constant value for the Lundquist number.
    Such models feature a more refined level of turbulence with respect to the case of intermediate constant diffusivity, consistent with the fact that the resistivity in the jet varies in the range $10^{-5}<\eta<10^{-3}$.
    However the injected energy saturates at different levels and at different times, because of the spatial and temporal variations of the resistivity.
    In our opinion the recipe based on a constant Lundquist number, that is $\eta = v_A L/S$ ($L$ is a typical scale, here the injection width, and $v_A$ is the Alfvén speed), is to be preferred for both computational reasons (it is not subject to the numerical diffusion of an advected passive scalar) and physical ones (the resistivity is basically proportional to the local magnetization), so we recommend its use.

\end{enumerate}

In conclusion, let us briefly discuss some important physical implications and possible lines of future development of the present work.

As illustrated by our simulations, the importance of addressing the large scale-separation between jet and fundamental plasma dynamics cannot be overstated.
The use of high-accuracy numerical schemes is thus crucial to quantitatively  assess the impact of turbulent diffusion in relativistic jets.
Future work will therefore need to improve upon this aspect, e.g. by considering global high-order schemes \citep{Berta2023}.

Another debated issue is the possible role played by the \textit{resistive} electric field, that is the non-ideal component proportional to the current density in the simplest Ohmic-type modelization, especially for the acceleration of the highest energy particles (see, e.g., \citealt{Kowaletal2009,Medetal2021,Sironi2022,PMB2022} and references therein).
A possible scenario arising from PIC simulations is that the non-ideal field (at X-points) plays an important role in non-thermal particle acceleration in microscopic current sheets as it starts the acceleration process. 
After this, the Fermi-type acceleration via the ideal electric field in the evolving plasmoids takes over, and most of the particle energy is gained \citep[e.g.][]{Guo2019}. 
At the much larger macroscopic scale of astrophysical jets, further acceleration is likely to take place via the ideal electric field of MHD turbulence \citep[e.g.][]{Medetal2023}.
This large-scale Fermi acceleration can be studied using test particles or PIC-MHD methods \citep{MBVM2018,VMBRM2018,Medetal2023}, even in the context of propagating relativistic jets. 

Due to the large separation in scales, the precise interplay between the microscopic ones (i.e. the current sheets typically studied via PIC simulations) and the large scales typical of MHD simulations of astrophysical sources is still very unclear.
However, we would like to stress here that since the real resistivity is orders of magnitude lower than the values adopted in this paper (and the numerical diffusivity typical of any ideal RMHD simulation), the latter should not be strictly interpreted as the microscopic resistivity coming from the PIC simulations, but rather as an effective "sub-grid" resistivity, attempting to model the impact of unresolved and non-ideal kinetic processes on the larger scales.
The effectiveness of such sub-grid models (which in several cases combine turbulent diffusion and mean-field dynamo action) has been positively tested in different astrophysical contexts by means of high-resolution studies \citep[e.g.][]{Liska2020,TDzBB2020,Guilet2022,Reboul-salze2022,Kiuchi2023}.
More refined resistivity models (see, e.g. \citealt{SPRBSK2023}) will be able to mimic with even better consistency the impact of microscopic kinetic process on the large scale scenario.


Other than AGN and blazar jets, the type of resistive simulations described here could also apply to different type of sources. A noticeable example is provided by the short-GRB jets arising from BNS merger multimessenger events \citep{PCKM2021,PCVM2022}, where the magnetic field is known to play a crucial role \citep{Ciolfi2020}.
For relativistic jets propagating in such a complex and structured magnetized environment, a finite conductivity is expected to be important and to lead to very different observational signatures, so this will certainly be a direction for future investigation.

\section*{Acknowledgements}

We thank Christian Fendt for useful comments/discussions.
LDZ, AP, RC, and AM acknowledge support from the PRIN-INAF 2019 Grant “Short gamma-ray burst jets from binary neutron star mergers” (Ob.\,fu.\,1.05.01.85.16).
All the simulations were performed on GALILEO100 and MARCONI100 clusters at CINECA (Bologna, Italy). In particular, we acknowledge CINECA for the availability of high performance computing resources and support through an award under the MoU INAF–CINECA (Grant INA23\_C9B05) and through a CINECA–INFN agreement, providing the allocations INF22\_teongrav and INF23\_teongrav.
This project has received funding from the European Union's Horizon Europe research and innovation programme under the Marie Sk\l{}odowska-Curie grant agreement No 101064953 (GR-PLUTO), and by ICSC – Centro Nazionale di Ricerca in High Performance Computing, Big Data and Quantum Computing, funded by European Union – NextGenerationEU.

\section*{Data availability}

The PLUTO code is publicly available. Simulation data will be shared on reasonable request to the corresponding author.

\bibliographystyle{aa.bst}
\bibliography{paper}

\appendix
\section{Numerical benchmarks}
\label{App::EosTests}

The implementation in the PLUTO code of the IMEX scheme coupled to the Taub Eos (see Section \ref{sec:IMEX-EoS}) has been tested through a series of test problems, which we report here. In particular, we show two 1D shock tubes (for which the impact of the EoS has been described in \citealt{MigMck2007}) and a 2D Kelvin Helmholtz instability numerical benchmarks, employing Cartesian coordinates.
Since, to the extent of our knowledge, the exact solution of the Riemann problem of the RRMHD equations has not been recovered yet, for the shock tube tests we focused on the regimes where the solution approaches the ideal one ($\eta\to0$) and where the solution approaches the frozen limit ($\eta\to\infty$).
In the frozen limit, the hydrodynamic quantities are decoupled from the electromagnetic ones \citep{MMB2018}; therefore, the choice of the EoS should not affect the solution for the electric and magnetic fields, while the fluid variables obviously evolve differently.
Unless stated otherwise, we adopt the same numerical algorithms used for the main simulations.

\subsection{Shock tube n. 1 (ST1)}
\label{sec:ST1}
%

The first shock tube problem (ST1), performed previously by \citet{Balsara2001,DZBL2003}, represents a 1D explosion driven by an initial pressure difference (a factor 30 in thermal pressure and 100 in magnetic pressure).
We choose a domain $x\in[-0.5,0.5]$ (therefore the discontinuity is placed at $x = 0$) and a resolution of $N_x = 400$ grid cells.
The initial conditions for the discontinuous quantities are:
\begin{equation}
(p,\,B_y,\,B_z) = \left\{
\begin{array}{ll}
(30.0,\,6.0,\,6.0) \qquad & x <    0, \\ \noalign{\medskip}
(1.0,\,0.7,\,0.7)\qquad   & x \geq 0,
\end{array}\right.
\end{equation}
while we set $\rho = 1.0$, $B_x = 5.0$, and $\vec{v} = \vec{E} = 0$ throughout the entire domain.
We performed two runs with constant uniform resistivity with, respectively, $\eta = 10^{-6}$ (to reproduce the ideal limit) and $\eta = 10^{40}$ (to reproduce the frozen limit).
The final time of each simulation is $t = 0.4$.
As pointed out by \citet{Balsara2001}, the emerging wave pattern (in the ideal regime) consists in a contact discontinuity separating two (fast and slow) left-going rarefaction waves from two (fast and slow) right-going shocks.
In the frozen limit the solution consists of a similar wave pattern; however, the fast and slow waves are replaced by, respectively, light and sound waves.
No rotational discontinuity is seen.

The shock tube results are compared with two reference runs: one, computed by solving the ideal RMHD equations with the Taub Eos, and one computed by solving the resistive RMHD equations in the frozen limit with the ideal EoS ($\Gamma = 5/3$).
Both reference runs are performed by employing the same numerical algorithms stated before (note that, since the ideal RMHD equations do not have a stiff component, the time integration is computed through a $3^{\rm rd}$ Runge-Kutta time integrator) and a resolution of $N_x = 12800$.

\begin{figure}
    \centering
    \includegraphics[width=0.49\textwidth]{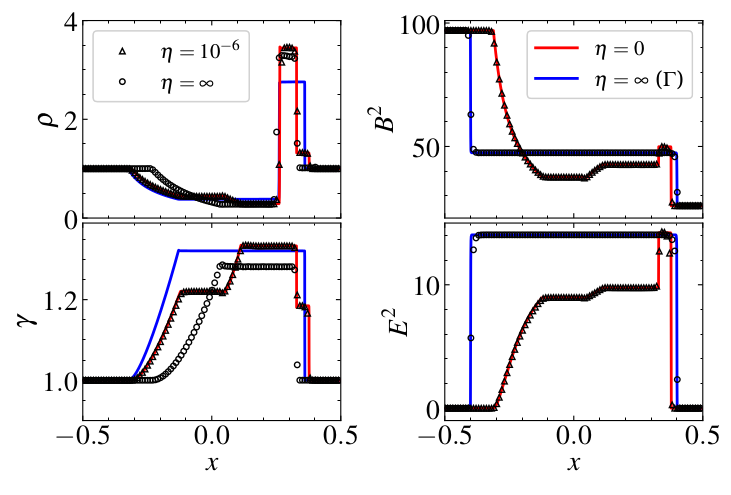}
    \caption{ST1: comparison of fluid and electromagnetic quantities. On the left panels are shown the fluid density (top) and the Lorentz factor (bottom), while on the right panel the squared norm of the magnetic (top) and electric (bottom) fields are is shown.
    The solid lines represent the reference simulation respectively in the ideal (red) and frozen (blue) limits, while the test simulations are shown through triangles (ideal limit) and circles (frozen limit).}
    \label{fig::st1_eos}
\end{figure}

In Fig. \ref{fig::st1_eos} we show, density and Lorentz factor in the left (respectively top and bottom) panels, while the square of magnetic and electric field are shown in the right (respectively top and bottom) panels.
For sake of readability, the solid lines (which represent the reference simulations, respectively the ideal regime in red and the frozen limit with ideal EoS in blue) are superimposed by a fraction of the data points ($1/4$, i.e. 100 data points are shown) computed through our novel scheme (respectively, triangles for the ideal regime and circles for the frozen limit).
As expected, no impact of the EoS is found on $E^2$ and $B^2$ in the frozen limits, as well as both the electromagnetic field and the fluid variables in the ideal limit.
Note that, in the frozen limit, the sound and the light waves are split between the fluid and the electromagnetic variables. As a result, in the frozen limit the wave pattern is always composed by two waves for each variable (plus the contact wave for the density). 

\subsection{Shock tube n. 2 (ST2)}
\label{sec:ST2}
%

\begin{figure}
    \centering
    \includegraphics[width=0.49\textwidth]{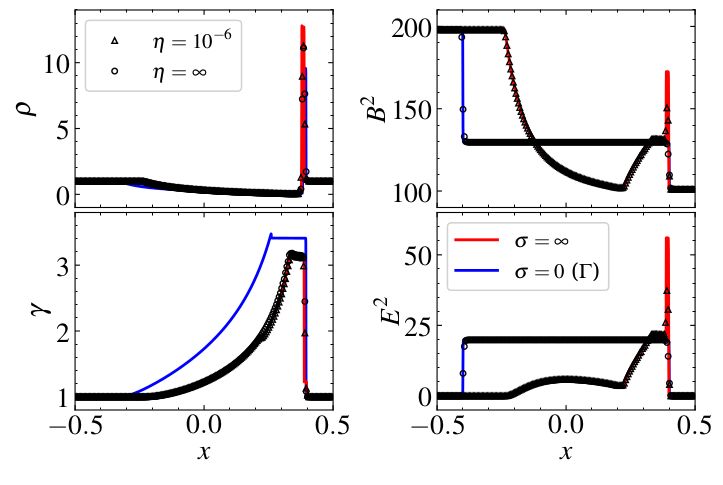}
    \caption{Same quantities as in Fig. \ref{fig::st1_eos}, here for the ST2 test.}
    \label{fig::st2_eos}
\end{figure}

\begin{figure*}
    \centering
    \includegraphics[width=0.87\textwidth]{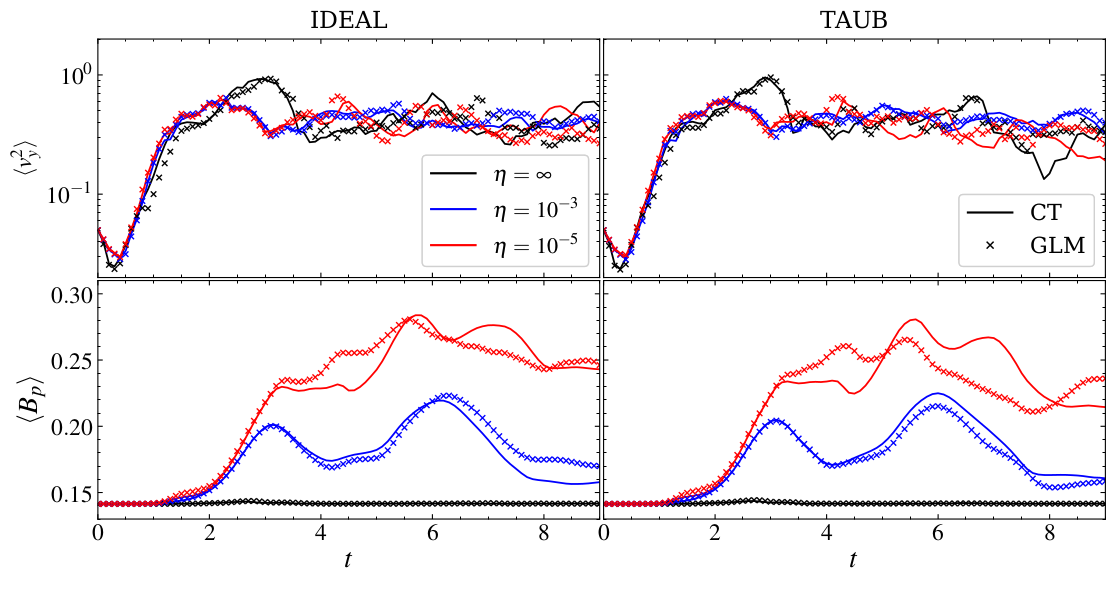}
    \caption{Kelvin Helmholtz instability. On the left/right panels are plotted the growth rate (top) and the poloidal magnetic field (bottom) are shown with ideal/Taub EoS.}
    \label{fig::khi_gr}
\end{figure*}

The second shock tube problem, again performed by \citet{Balsara2001,DZBL2003}, consists of a stronger 1D blast wave (here the jump in thermal pressure is of a factor as high as $10^4$), which yields a more relativistic configuration.
The setup is very similar to the previous benchmark; in fact, the only change is in the initial conditions, where discontinuous quantities are initialized as:
\begin{equation}
(p,\,B_y,\,B_z) = \left\{
\begin{array}{ll}
(1000.0,\,7.0,\,7.0) \qquad & x <    0, \\ \noalign{\medskip}
(0.1,\,0.7,\,0.7)\qquad   & x \geq 0,
\end{array}\right.
\end{equation}
while we set $\rho =1.0$, $B_x = 10.0$, and $\vec{v} = \vec{E} = 0$.
The results are shown in the same fashion of the previous benchmark.
The solution is composed of two left (fast and slow) rarefaction waves and two right (fast and slow) shocks in the ideal regime, and of two light and sound (left and right) waves in the frozen limit.
Again, no rotational discontinuity is shown.

Both the frozen and the ideal limits are reproduced with great degree of accuracy, as shown in Fig. \ref{fig::st2_eos}.
The sharp discontinuities located at $x\approx0.4$ are captured in both regimes despite the lower resolution.
As pointed out in \citet{MigMck2007}, the sound speed is highly overestimated while applying an ideal EoS.
Such behavior can be clearly seen in the frozen limits, where the density and the Lorentz factor feature discontinuities propagating at the sound speed (as shown by the circles and the blue solid line).

\subsection{Kelvin-Helmholtz instability (KHI)}
\label{sec:KHI}
%

In order to compare the different equations of state and to validate the numerical algorithms adopted in the paper, we performed the Kelvin-Helmholtz instability test of \citet{Mizuno2013}.
The domain consists of a Cartesian box with $x\in[-0.5,0.5]$ and $y\in[-1.0,1.0]$, and is filled with an initial constant gas pressure $p = 1.0$.
The shear velocity profile is given by:
\begin{equation}
v_x = v_{\rm sh}\tanh\left(\frac{|y| - 0.5}{a}\right)= v_{\rm sh}\tanh\phi(y) ,
\end{equation}
where $v_{\rm sh} = 0.5$ and $a = 0.01$ are, respectively, the maximum shear velocity and the characteristic thickness of the double shear layer.
The initial density is set to:
\begin{equation}
\rho = \frac{1 + \tanh\phi(y)}{2}\rho_h + \frac{1 - \tanh\phi(y)}{2}\rho_l,
\end{equation}
where $\rho_h = 1$ and $\rho_l = 10^{-2}$.
The magnetic field is provided in terms of the plasma-$\beta$:
\begin{equation}
    (B_x,\,B_y,\,B_z) = \left(\sqrt{\frac{2p}{\beta_p}},\,0,\,\sqrt{\frac{2p}{\beta_t}}\right).
\end{equation}
The equilibrium is perturbed by adding:
\begin{equation}
 v_y = \sign(y)A_0v_{\rm sh}\sin(2\pi x)\exp\left(-\frac{\phi(y)^2a^2}{\alpha^2}\right)
\end{equation}
where the perturbation is characterized by the amplitude $A_0 = 0.1$ and the characteristic length scale of decay $\alpha = 0.1$.
Lastly, we set $v_z = 0$ and we compute the electric field from its ideal condition.
The final time of the simulation is $t = 15$.

In Fig. \ref{fig::khi_gr} we show the growth rate, defined as the volume average of $v_y^2$ (top panels) and the growth of the volume-averaged poloidal field (bottom panels) using the ideal (left panels) and the Taub (right panels) equations of state, and the UCT (solid lines) and GLM (x-symbols) algorithms to preserve the non-evolutionary Maxwell equations.

We find good agreement with the values obtained by \citet{Mizuno2013,MMBZ2019}; as expected, the growth rate does not show any dependence on the resistivity, regardless of the equation of state.
Conversely, a higher resistivity is able to quench the poloidal magnetic field amplification, since the latter is caused by the stretching due to the main vortex formation.
Moreover, the ideal EoS leads to a stronger amplification of the poloidal magnetic field, in agreement with \citet{Mizuno2013}.
We also notice very good agreement between the UCT and the GLM algorithms until $t = 3$, i.e. after the non-linear regime has already taken place.

\section{Turbulence level and resolution}
\label{App::turbulence}

In order to quantify the agreement in the level of turbulence among different models in the paper, we have computed the effective resolution of our simulations by computing the normalized gradient $\langle|\nabla Q|/Q\rangle$
of the density $\rho$ and the magnetic energy $B^2/2$, similarly to \citet{MUB2009,BecSto2011}.

\begin{figure}
    \centering
    \includegraphics[width=0.45\textwidth]{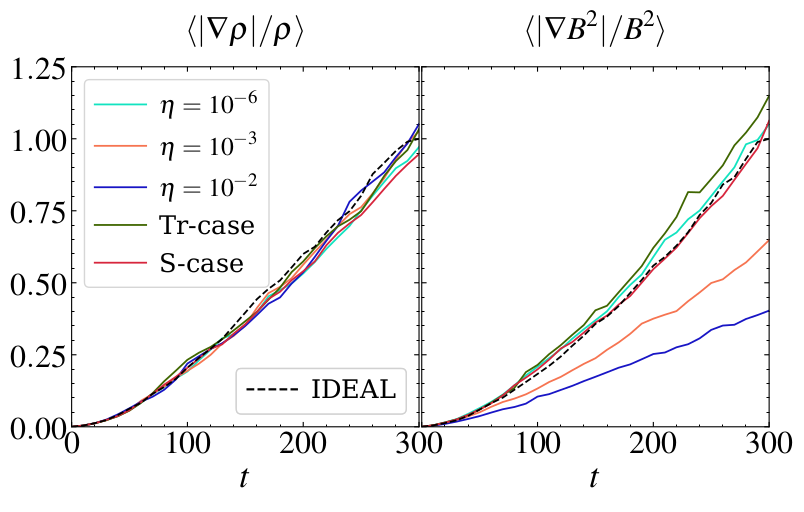}
    \caption{Effective turbulence resolution (computed as $\langle|\nabla Q|/Q\rangle$) of density (left panel) and magnetic energy (right panel) computed for different resistivity models and the ideal simulation using the same setup as \citet{MUB2009}.}
    \label{fig::turbulence}
\end{figure}

\begin{figure}
    \centering
    \includegraphics[width=0.45\textwidth]{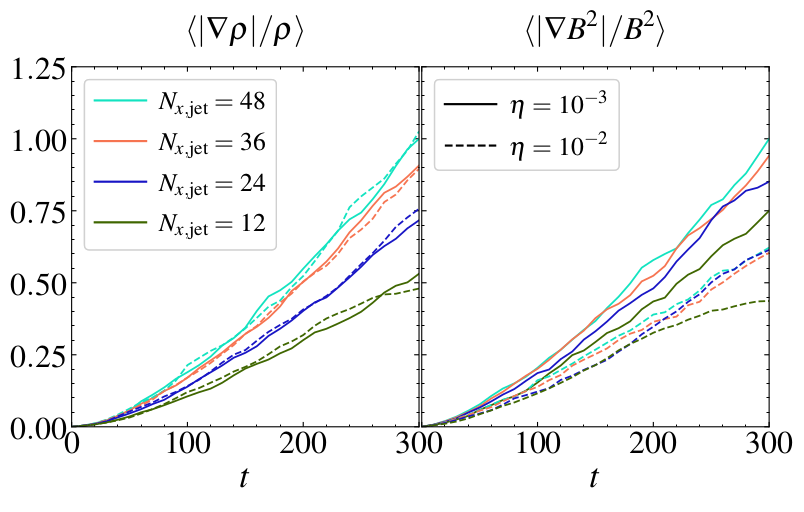}
    \caption{Same of Fig. \ref{fig::turbulence} but for different resolution and constant resistivity $\eta = 10^{-3}$ (solid line) and $\eta = 10^{-2}$ (dashed line).}
    \label{fig::resolution}
\end{figure}

In Fig, \ref{fig::turbulence} we show a comparison between all the runs performed in the paper and a reference simulation computed in the ideal RMHD regime and using the same numerical setup presented in \citet{MUB2009} (but with resolution matching the one used for our models).
All the quantities are normalized to the ones at the final time of the ideal simulation.
As expected, all the runs show excellent agreement (i.e. discrepancies are $\lesssim15\%$) in density (left panel) regardless of the specific value of $\eta$, while the magnetic field turbulence is affected by a sufficiently strong physical resistivity.
In particular, we notice that the low-resistivity case, as well of the non-constant diffusivity runs show a turbulence which is comparable with the ideal simulation, while the runs with higher resistivity show a lower level of turbulence (respectively, $65\%$ and $40\%$ for the cases $\eta = 10^{-3}$ and $\eta = 10^{-2}$).

On the other hand, the turbulent scale required a sufficiently high resolution in order to be properly resolved.
A "poor" resolution would lead to a higher numerical diffusivity which would strongly affect the turbulent scale and therefore the dynamical evolution of the jet.
In order to prove that our resolution is enough to capture the magnetic dissipation scales dictated by the physical resistivity, we compared the normalized gradients for the case $\eta = 10^{-3}$ and $\eta = 10^{-2}$ by adopting different grid resolutions.
As shown in Fig. \ref{fig::resolution}, the magnetic turbulence (right panel) shows very good agreement even at half (i.e. 24 cells per $r_j$) of the reference resolution, while the poorest resolution model (i.e. 12 cells per $r_j$) shows significant qualitative differences.
Concerning the density turbulence, despite slight differences between 36 and 48 cells per $r_j$ ($\lesssim10\%$), we find that our reference resolution is high enough to reach convergence.
Lower grid resolutions, on the other hand, lead to a more suppressed fluid turbulence.

\end{document}